\documentclass{emulateapj}

\usepackage[latin1]{inputenc}
\usepackage[T1]{fontenc}
\usepackage[english]{babel}
\usepackage{amssymb}
\usepackage[]{natbib}
\usepackage{rotating}

\begin{document}

\slugcomment{Accepted for publication in ApJ}

\title{N$_2$H$^+$ and N$_2$D$^+$ in Interstellar Molecular Clouds.
II -- Observations}
\author{F. Daniel\altaffilmark{1,2},
J. Cernicharo\altaffilmark{1},
E. Roueff\altaffilmark{3},
M. Gerin\altaffilmark{4},
and M.L Dubernet\altaffilmark{2}}
\email{daniel@damir.iem.csic.es, cerni@damir.iem.csic.es,
marie--lise.dubernet@obspm.fr}
\altaffiltext{1}{Dept. Molecular and Infrared Astrophysics (DAMIR),
Consejo Superior de Investigaciones Cient\'{\i}ficas (CSIC),
C/ Serrano 121, 28006 Madrid. Spain}
\altaffiltext{2}{Observatoire de Paris--Meudon, LERMA UMR 8112, 5 Place
Jules Janssen, F--92195 Meudon Cedex, France.}
\altaffiltext{3}{Observatoire de Paris--Meudon, LUTH  UMR 8102, 5 Place
Jules Janssen, F--92195 Meudon Cedex, France.}
\altaffiltext{4}{Observatoire de Paris--Meudon, UMR 8112}

\defcitealias{dan06}{Paper I}

\begin{abstract}
We present observations of the $J$=1--0, 2--1, and 3--2 rotational transitions of
N$_2$H$^+$ and N$_2$D$^+$ towards a sample of prototypical dark clouds.
The data have been interpreted using non--local radiative transfer
models. For all sources previously studied through millimeter continuum observations,
we find a good agreement between the volume density estimated
from our N$_2$H$^+$ data and that estimated from the dust emission.
This confirms that N$_2$H$^+$ depletion is not very efficient in dark clouds
for densities as large as 10$^6$ cm$^{-3}$, and also points out that
a simultaneous analysis based on mm continuum, N$_2$H$^+$ and N$_2$D$^+$
observations should lead to reliable estimates for the temperature and 
density structure of cold dark clouds. 

From multi--line modeling of N$_2$H$^+$ and N$_2$D$^+$, we derive the 
deuterium enrichment in the observed clouds. Our estimates are 
similar or higher than previous ones. The differences
can be explained by the assumptions made on the cloud density profile,
and by the chemical fractionation occurring in the clouds.
For two of the observed objects, L183 and TMC2, multi--position observations
have allowed us to derive the variation of the N$_2$D$^+$/N$_2$H$^+$
abundance ratio with the radius. We have found that it decreases by an
order of magnitude for radii greater than a few 0.01 pc (i.e. outside
the central cores). Inside the dense condensations, the fractionation is
efficient and, compared to the abundance ratio expected from statistical 
considerations based on the cosmic D/H ratio, the deuterium enrichment 
is estimated to be $\simeq$ 0.1--0.5 10$^5$.

An important result from our observations and models is that
the interpretation of deuterated molecular species emerging intensities
in cold dark clouds requires specific radiative transfer 
modeling because
the excitation conditions for the deuterated species can be 
quite different from those of the main isotopologue. Moreover, the use 
of three rotational lines for N$_2$H$^+$ and N$_2$D$^+$
allows to constrain the size of the emitting regions for each species and 
to determine accurately the volume density. This enables to draw a 
detailed picture of the spatial variation of deuterium enrichment.

\end{abstract}

\section{Introduction}
Over the past decades, important efforts have been devoted to improve
our understanding of the physical conditions in protostellar clouds. 
These studies have enabled to draw an evolutionary
sequence of the stages prior to
the formation of protostars, through the morphological, dynamical and
chemical characteristics
of the clouds. Recent progress in detection capabilities in the far 
infrared, sub--mm and mm windows, have permitted to constrain
the density distribution of molecular clouds through measurements of dust
absorption and emission \citep[see e.g.][]{war94,war99}. It has been found that 
the geometrical structure of cold dark clouds is consistent
with an inner region of nearly constant density and a 
surrounding envelope that has a density which decreases as a power law. 
The density distribution thus obtained is often
used as a starting point for molecular line studies and allows to 
constrain the spatial distribution of the molecules \citep[see e.g.][]{taf04}. 
Moreover, from NH$_3$ inversion lines it has been found that the clouds are 
nearly isothermal \citep{ben89,taf04}. Such observational constraints are of great interest 
to understand the physical mechanisms that drive the collapse. 
However, the balance between gravitational energy 
and support mechanisms such as thermal pressure, turbulence or 
magnetism is yet to be understood \citep{aik05}. Moreover, theoretical studies 
predict that the initial conditions strongly affect the way 
the collapse of the cloud evolves. In particular, time 
dependent studies, which couple both dynamical and chemical processes, have shown
that molecules can be used as tools to probe the stage of evolution prior 
to the formation of protostars. These studies also predict that deuterium
fractionation increases with time and that the D/H ratio can reach high values 
for a variety of species \citep{rob03}. 

The goal of this work is to put observational constraints on the
spatial structure of a sample of dark clouds through
observations of a late type molecule, N$_2$H$^+$, and of its deuterated 
isotopologue N$_2$D$^+$. These molecules are well adapted to probe the 
innermost dense regions of dark clouds due to their high dipole moments and also
because they are weakly depleted onto dust grains. 
While molecular observations are a very powerful tool to derive the 
physical and chemical conditions of the gas in these regions, the 
interpretation of the observations needs essential informations on molecular 
physics data, such as collisional rate coefficients, line intensities and frequencies.
Molecules often used as tools, like N$_2$H$^+$, have been
interpreted so far with crude estimates of these parameters. Recently, \citet{dan05} 
have computed collisional rate coefficients between N$_2$H$^+$ and He for 
the range of temperatures prevailing in cold dark clouds and low mass star 
forming regions. \citet{dan06} (hereafter referred to as \citetalias{dan06}) 
have modeled, using these new rates, the emission of N$_2$H$^+$ in dark clouds.
A wide range of physical conditions has been explored in order to
have a general view of the different processes
that could lead to the observed intensities of N$_2$H$^+$ hyperfine lines.

Observations are described in section \ref{observations}.
In section \ref{clouds} we present the 
results obtained for each cloud and we derive the physical
conditions of the gas as well as N$_2$H$^+$ and N$_2$D$^+$ abundances.
The conclusions are given in section \ref{conclusion}.
The excitation processes occurring for deuterated species are analyzed
in the Appendix.

\section{Observations}\label{observations}

N$_2$H$^+$ and  N$_2$D$^+$ observations were performed during several 
observing sessions. Additionally, we make use of N$_2$D$^+$ data already 
discussed by \citet{tin00}. 

Observations of the N$_2$D$^+$ $J$=2--1 line have been
obtained at the IRAM--30m telescope (Granada, Spain) in July 2000 during average
weather conditions. The pointing was
checked regularly on nearby continuum sources, and the focus was
checked on small planets.
The data have been taken in frequency switching mode, with the mixers
connected to the versatile, high spectral resolution, correlator VESPA.
The N$_2$D$^+$ $J$=1--0 data have been taken in August 2004 when the tuning range
of the 3mm mixers was extended to frequencies lower than 80 GHz. Simultaneous
observations of N$_2$D$^+$ $J$=3--2 were performed using two mixers tuned
at this frequency. The system temperature was $\sim 250$ K at 77 GHz
and $\sim 500$ K at 231 GHz. Particular care was given to the intensity 
calibration as the mixer tuning 
is done without side band rejection at these low frequencies. The
image side band gain has been measured on an offset position
or a continuum source for each object, and applied during the data processing.

Further observations of the $J$=1--0, 2--1 and 3--2 lines of N$_2$H$^+$ and of 
the N$_2$D$^+$ $J$=3--2 line were
carried out in January 2005 with the IRAM--30m telescope.
Four receivers tuned at the frequency of these
lines were used simultaneously. 
Pointing was checked every hour on continuum sources
close to the targets. The relative alignment of the receivers was checked
with Saturn and Jupiter and found to be better than 2.0''. Our observations
indicate that the offsets on the sky from the pointing derived with
the 3 mm receiver were 1.0'', 2.0'' and 1.0'' for the 2 mm, 1 mm low frequency
and 1 mm high frequency receivers respectively.
Focus was checked at the beginning and end of each observing shift. It
was found to be stable for the entire observing run.

The weather was excellent during observations with a typical water vapor amount above
the telescope site of $\simeq$ 1.0 mm, or even better.
Calibration was performed using two absorbers at ambient and liquid nitrogen
temperatures. The ATM package \citep{cer85a,par01} was
used to estimate the significant atmospheric parameters for the observations. 
The typical system temperatures were 100 K, 1000 K and 250 K at 3, 2 and 1 mm respectively.
The $J$=2--1 line was observed in upper side band which avoided the possibility
of rejecting the image side band in the 30--m receivers. The large
system temperature for this transition is due to the high opacity of
the atmosphere at 186 GHz and to the fact that the frequency is at
the edge of the tuning range of the receiver. Calibration uncertainties for 
the $J$=1--0 and 3--2 lines are well below 10\% (the standard calibration
accuracy under good weather conditions). However, the $J$=2--1 line has a larger
uncertainty due the effects commented above, and, in order to have an estimate
of the calibration accuracy we observed the continuum emission of
Jupiter, Saturn and Mars at different elevations. We have found that the
intensity of the $J$=2--1 transition was always compatible with the expected one. 
However, these double side band observations of the planets are not very sensitive
to the side band rejection. Nevertheless, these data allow us to derive beam 
efficiencies for each frequency. They are in excellent agreement with 
those published for the 30--m IRAM telescope. For some strong sources we 
observed several times a reference position and we noted that the $J$=2--1 line 
intensity had calibration errors up to 20\% depending on the elevation of the source.
We conclude that with the configuration adopted for the 2 mm receiver, the upper 
signal side band has a gain of 0.35 rather than 0.5 (perfect DSB).
This fact introduces systematic errors in the determination of the sky opacities. 
It is impossible to estimate the actual error for each source. Hence, for 
the $J$=2--1 line we assume a typical calibration error of 20\%. 
Nevertheless, most observed $J$=2--1 intensities agree with those expected 
from the results of the $J$=1--0 and 3--2 lines (see Sect. \ref{clouds}).
Only for TMC2, L1489 and L1251C it has been necessary to correct 
the $J$=2--1 intensities by multiplicative factors of, respectively, 0.8, 0.8 and 0.85  
in order to have data consistent with the well calibrated $J$=1--0 and
3--2 lines. 

\begin{figure}
\centering
\includegraphics[scale=0.65,angle=270]{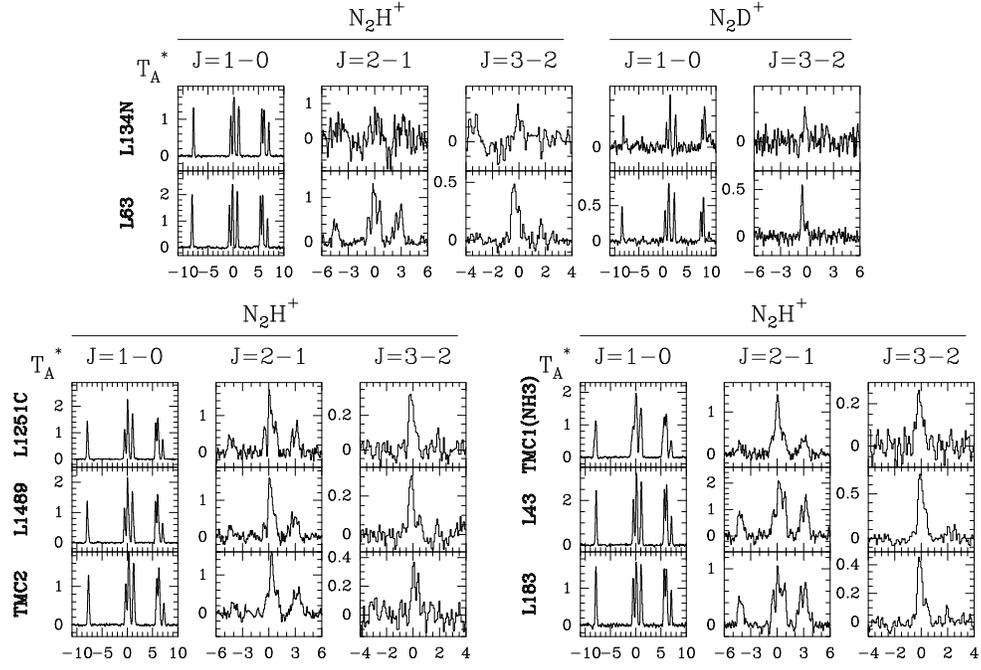}
\caption{Observed $J$=1--0, 2--1, and 3--2 transitions of N$_2$H$^+$ in
the direction of selected sources. Ordinate is antenna temperature
and abscissa is the V$_{LSR}$. For each transition the assigned frequency
corresponds to the one of the strongest hyperfine component.
  \label{graph:DATA}}
\end{figure}

The observations were performed in frequency switching mode. As all sources
have narrow lines, the baseline resulting from folding the
data has been removed by fitting a polynomial of degree 2.
Figure \ref{graph:DATA} shows the  N$_2$H$^+$ and  N$_2$D$^+$ observed
lines in selected
sources and Table \ref{table:obs_mod} gives the position of the clouds
observed in this paper
(see section \ref{clouds}). The intensity scale is antenna temperature, T$_A^*$. In order
to compare the results of our models with observations,
we have to convolve model results with the beam pattern of the 30--m IRAM telescope. 
We have adopted half power beam widths (HPBW) of 27'', 13.5''
and 9'', beam efficiencies of 0.76, 0.59 and 0.42,
and error beams of 350'', 220'', and 160'' for the
$J$=1--0, 2--1 and 3--2 lines of N$_2$H$^+$.
For N$_2$D$^+$, we adopted HPBW's of 32'', 17'' and 10.5'', beam
efficiencies of 0.79, 0.69 and 0.52 and the same error beams as for 
N$_2$H$^+$ lines.
The whole data set for all observed sources is shown 
in Figures 3 to 15.

\section{Results}\label{clouds}
Over the past decades, mm continuum observations and star count analysis \citep{war94}
have revealed that the density structure of cold dense cores 
closely matches a Bonnor--Ebert sphere, with an inner region of nearly uniform
density and steep outer edges where the density decreases as
r$^{-p}$ with p = 2 -- 2.5. This description is valid for
r $<$ 0.1 pc typically, and at larger scales (0.1 -- 1 pc), the density
structure of the outer envelope has been shown to vary 
from sources to sources. For example, \citet{cer85b} have found that a 
shallow density profile, n(r) $\propto$ r$^{-1/3}$, is adequate to model the 
density structure of the envelope of the Taurus--Auriga complex. 
In this section, we use a density profile which corresponds to n(r) = n$_0$ for r $<$ r$_0$ and 
n(r) = n$_0$(r$_0$/r)$^2$ for r $>$ r$_0$. In order to characterize the clouds, we determine 
the value of the central density, i.e. n$_0$, and the radius of the inner flat region, i.e. r$_0$. 
For L1517B, we compare the results obtained using this density profile with the  
phenomenological density law used in \citet{taf04}.  

The observed line profiles for N$_2$H$^+$ and N$_2$D$^+$ $J$=1--0, 2--1, and
3--2 lines are shown in Figures 3 to 15. In order to model our data we
have found that some additional constraints were necessary. In particular
the extent of the emission has been found to be a critical parameter to
derive physical conditions. Our data are in most cases related to
one single observed position (except for L183 for which we have a map in
several transitions). Fortunately, some clouds presented in this section have 
been mapped in the $J$=1--0 transitions
of N$_2$H$^+$ by \citet{cas02}, \citet{taf04}, and \citet{cra05}
(In the discussion of the sources, if not specified, N$_2$H$^+$ maps refer 
to data from \citet{cas02}). For six sources (L63, L43, L1489, L1498, 
L1517B, and TMC--2) they have reported $J$=1--0 integrated intensity versus impact parameter.
Thus, we have checked that the profile of the $J$=1--0 integrated intensity 
provided by our models was consistent with those data. 
For the other clouds, similar data could not be found in the literature.

The radiative transfer models aim at reproducing our $J$=1--0, 2--1 and 3--2
line intensities and the $J$=1--0 radial emission profile. A discussion concerning
the assumptions made to model N$_2$D$^+$ is given in
the Appendix (that deals in particular with the determination of collisional rate coefficients
for the isotopologue of N$_2$H$^+$).
Observed positions, central line intensities and adopted distances to 
the sources are given in Table \ref{table:obs_mod}. We report in the same table 
the $J$=1--0 emission half power radius (noted $r$). This corresponds to the radius 
which encloses half of the $J$=1--0 total intensity which emerges of the cloud.
The convolution of the emerging line intensities 
with the telescope beam pattern has been done taking 
into account the characteristics of 30--m IRAM radio telescope (see section 2)
and those of the 14--m FCRAO antenna. For the latter we have used a HPBW
of 54'' at 91 GHz and a beam efficiency of 0.51. Nevertheless, the error beam
size for the FCRAO antenna is not known but must be taken into account.
In our modeling, we arbitrarily assume a size of 180''.
As quoted above, the radial
intensity profile enables to constrain the way
density and abundance vary through the cloud.
We stress that the density profiles determined in the following section
depend essentially on the observations of the $J$=1--0 line. Hence,
an accurate description of the radiation pattern of the FCRAO antenna 
is needed to obtain the best estimates for the physical parameters of the clouds.
We have checked the effect of introducing different error beam sizes
for the FCRAO antenna, and the largest differences are obtained 
when the error beam is not taken
into account, which leads to $J$=1--0 line intensities 20--30\% lower
for the same physical conditions and source structure.
This effect could be compensated by increasing the size of the central
core, r$_0$, by 5''--10'', decreasing its density, n$_0$,
and increasing the abundance of N$_2$H$^+$ in the outer region of the
clouds.

For some sources, we have included infall
velocity fields in order to reproduce the spectral features seen in the $J$=2--1 line. 
The profiles used consist of a $ v \propto r $ function which 
corresponds to infalling cores for all 
sources, except for L1517B. Despite the inclusion of these velocity
fields, we have observed frequency offsets between the observations and the models.
In table \ref{table:VLSR}, we
report the V$_{LSR}$ determined from our $J$=1--0 data, the corrections to this
value for the other lines (V$_{LSR}$ and corrections are determined by a $\chi$
square fitting), and the velocity gradient found in the sources.
For all of them, we find that the $J$=2--1
and $J$=3--2 lines have V$_{LSR}$ $\sim$ 30 m s$^{-1}$ larger than
the V$_{LSR}$ of the $J$=1--0 line.
This would be consistent with the view of a static core with an 
infalling envelope as the $J$=2--1 and $J$=3--2 lines arise from the densest
part of the cloud. However, we think that the effect, 20 and 30 kHz 
for the $J$=2--1 and $J$=3--2 lines respectively, is within the 
accuracy of the laboratory frequencies of N$_2$H$^+$ 
\citep[see e.g.][]{ama05}. Consequently, the range of model parameters
is limited by the laboratory accuracy in the frequencies. 

\subsection{Individual core properties}

In this section, the cloud physical parameters (see Tables \ref{table:models} and \ref{table:models2}),
obtained from the modeling of our data, are discussed.
In the following discussion, we use the notation $r_D$ $\equiv$ X(N$_2$D$^+$)/X(N$_2$H$^+$)
to refer to the isotopologues abundance ratio. This quantity differs from the  
column density ratio (referred to as $R$ $\equiv$ N(N$_2$D$^+$)/N(N$_2$H$^+$)) in the case of three
clouds (L63, TMC2 and the two molecular peaks in L183) for which $r_D$ has been found 
to change with radius. R represents in all cases an averaged value along the line of 
sight of the N$_2$H$^+$ volume density.

For each cloud, the error bars quoted for the physical parameters are derived by use
of a reduced $\chi^2$. These are obtained by assuming that the parameters of the models are independent and thus, 
it does not account for the correlation among them. Nevertheless, it gives an indication on the sensitivity
of the emerging spectra to the model's parameters. The same criterion was used for both N$_2$H$^+$ and N$_2$D$^+$ spectra. 
The uncertainty on $r_D$ is then found important due to the lower signal to noise ratio of the N$_2$D$^+$ data and 
this leads to large error bars for the N$_2$D$^+$ column density (varying from 20\% to 100\% depending on the source)
while typically, the N$_2$H$^+$ column density is estimated with an accuracy of 10-20 \%.

\subsubsection{L1517B}

Modeled lines and observations of the $J$=1--0 and $J$=2--1 transitions of 
N$_2$H$^+$ and N$_2$D$^+$ are shown on figure \ref{graph:L1517B}.
When comparing the line profiles of the two isotopologues, it is important 
to note the differences which are introduced by their respective abundances, i.e.
opacities \citepalias[see][]{dan06}. 
We see that the intensity ratio between the different $J$=1--0 hyperfine components
are best reproduced for N$_2$D$^+$ which is optically thinner than N$_2$H$^+$.
This fact supports the idea that hyperfine intensity anomalies are due to 
radiative effects which occur at high opacities.
For the $J$=2--1 lines of the two isotopologues, we see that the main
central emission component is self absorbed for N$_2$H$^+$ while it is not 
for N$_2$D$^+$. Moreover, as commented in \citetalias{dan06}, the line profile seen
in the $J$=2--1 line of N$_2$H$^+$ is characteristic of expanding flows.

\begin{figure}
\centering
\includegraphics[scale=0.6,angle=270]{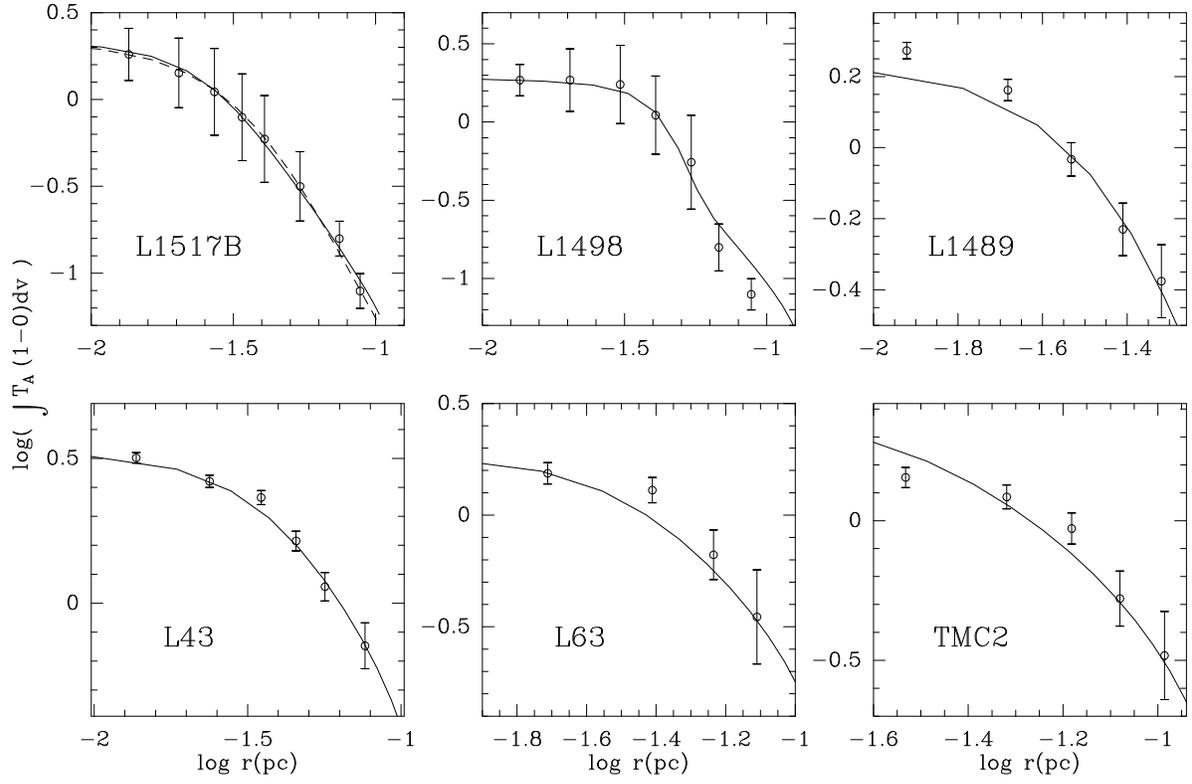}
\caption{ 
Observed radial integrated intensity profiles of L1517B, L1498,
L1489, L43, L63 and TMC2 compared to the models.
Observational data are from \citet{cas02} (obtained
with the FCRAO antenna) and \citet{taf04} for L1498 and L1517B 
(30m, Pico Veleta). For each source, the solid lines refer
to the emerging intensities of our modeling (see text).
For L1517B, the solid line correspond to 
the density profile used in this work and the dashed one to the 
profile used in \citet{taf04}.
\label{graph:int_rad}}
\end{figure}

\begin{figure}
\centering
\includegraphics[scale=0.7,angle=0]{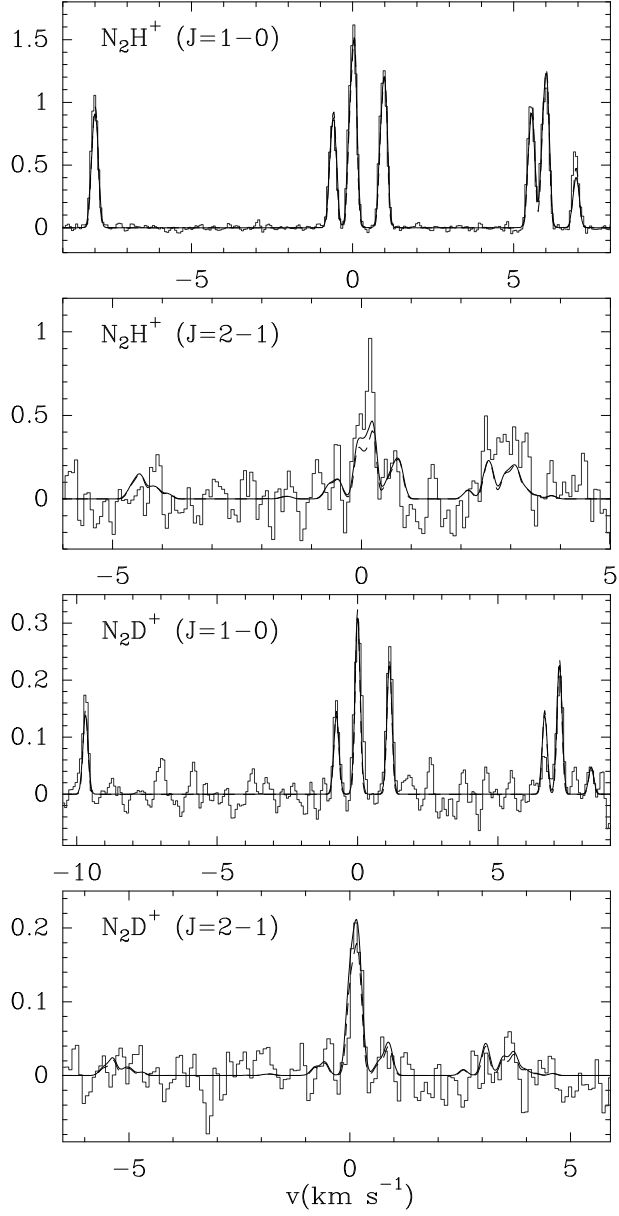}
\caption{ 
Observed and modeled $J$=1--0 and 2--1 transitions of N$_2$H$^+$
and N$_2$D$^+$ towards L1517B.
The solid line corresponds to the 
density profile used in this work and the dashed one corresponds
to the profile used in \citet{taf04}.
\label{graph:L1517B}}
\end{figure}

A detailed study of L1517B has been reported in \citet{taf04} where 
the structure of the cloud is derived from maps of the NH$_3$ inversion 
line and 1.2 mm continuum emission. The position we observed corresponds 
to the N$_2$H$^+$ peak intensity given by \citet{cas02},
which is situated at (+15'',+15'') of the N$_2$H$^+$ emission peak 
in the higher resolution map reported by \citet{taf04}. The latter 
position is also a maximum of the 1.2 mm emission map.
In our models we took into account the $J$=1--0 and $J$=3--2 lines 
reported by \citet{taf04} towards the continuum peak. 
On figure \ref{graph:int_rad}, we compare the $J$=1--0 radial
integrated intensity of our model to the observational data
reported in the latter study. 

To characterize the velocity structure of L1517B, we introduced
a linear outward gradient of 1.4 km s$^{-1}$ pc$^{-1}$ for radii beyond 20''.
As pointed out in \citetalias{dan06}, it is rather
difficult to constrain velocity fields from observations of the
$J$=1--0 and $J$=3--2 lines of N$_2$H$^+$, while the larger opacity of the $J$=2--1 line
enables a better estimate. 
From observations of CS emission lines, \citet{taf04} derived a value for the
outward gradient of 2 km s$^{-1}$ pc$^{-1}$, for radii larger than $\sim$ 70''. 
Hence, this velocity field is consistent with their observations of 
the $J$=1--0 and $J$=3--2 lines of N$_2$H$^+$ observed towards the central position.

On Figure \ref{graph:int_rad}, we compare the $J$=1--0 radial intensities 
obtained with two types of density profiles.
The first is the one quoted above (labeled  (1) in Table 
\ref{table:models}) and the second corresponds to the phenomenological 
profile used by \citet{taf04} (labeled as (2)). 
Using the profile (2), we obtain the same estimate than in the latter 
study, i.e. n$_0$ = 2.0 10$^5$ cm $^{-3}$ and r$_0$ = 35''. Moreover, we see 
that with the present observational data, we are rather insensitive 
to the exact form of the function which describes the 
density variations in the cloud, since the two density profiles 
give similar results for both line shapes and the $J$=1--0 integrated intensity 
profile. 

Using the two density profiles mentioned above, we also derive different estimates
for the N$_2$H$^+$ abundance, i.e. 1.6 10$^{-10}$ with (1) and 2.5 10$^{-10}$
with (2). In the latter case, our estimate is higher  
by $\sim$ 70\% with respect to the one obtained by \citet{taf04}, i.e. 
X(N$_2$H$^+$) = 1.5 10$^{-10}$. The difference is due to the use
of different sets of collisional rate coefficients in each studies.

We have determined the D/H ratio that allows to reproduce the $J$=2--1 line 
of N$_2$D$^+$ for the two density profiles. For 
the density profile (1), we derive R = 0.12$^{+0.04}_{-0.03}$ which is a factor 2 higher than 
the estimate of 0.06 $\pm$ 0.01 obtained by \citet{cra05} toward the density peak position.

\citet{ben89} have estimated a mass of 0.33 M$_{\sun}$ within a 
radius of 1.4', which corresponds to a mean volume density of 
n(H$_2$)=6 10$^3$ cm$^{-3}$. For the same radius, we obtain a mass 
of 2.5 M$_{\sun}$. On the other hand, the analysis of 
emission maps at 850 $\mu$m and 450 $\mu$m \citep{kir05} lead to an 
estimate of n(H$_2$) $\sim$ 4--5 10$^5$ cm$^{-3}$, in qualitative 
agreement with the central density assessed from our N$_2$H$^+$ data.

\subsubsection{L183}

L183 is a well studied object due to its proximity (d=110 pc). It has a non spherical shape
with a north--south axis in the density distribution. Recent studies point out a complex 
chemistry in the cloud, with four main molecular peaks showing characteristics which are interpreted 
as signs of different stages of evolution \citep{dic00}. The N$_2$H$^+$ 
maximum of intensity is situated near the densest 
part of the cloud (molecular peak C). A secondary maximum situated $\sim$ 3' to the north 
(molecular peak N) is also seen in the N$_2$H$^+$ intensity map.
For this source, the observed position for molecular peak C corresponds to the 
intensity peak as given by \citet{cas02}. It is placed at (-5'',+15'') with respect 
to the peak position of the higher resolution maps given by \citet{dic00} and \citet{pag05}.

\begin{figure}
\centering
\includegraphics[scale=0.7,angle=0]{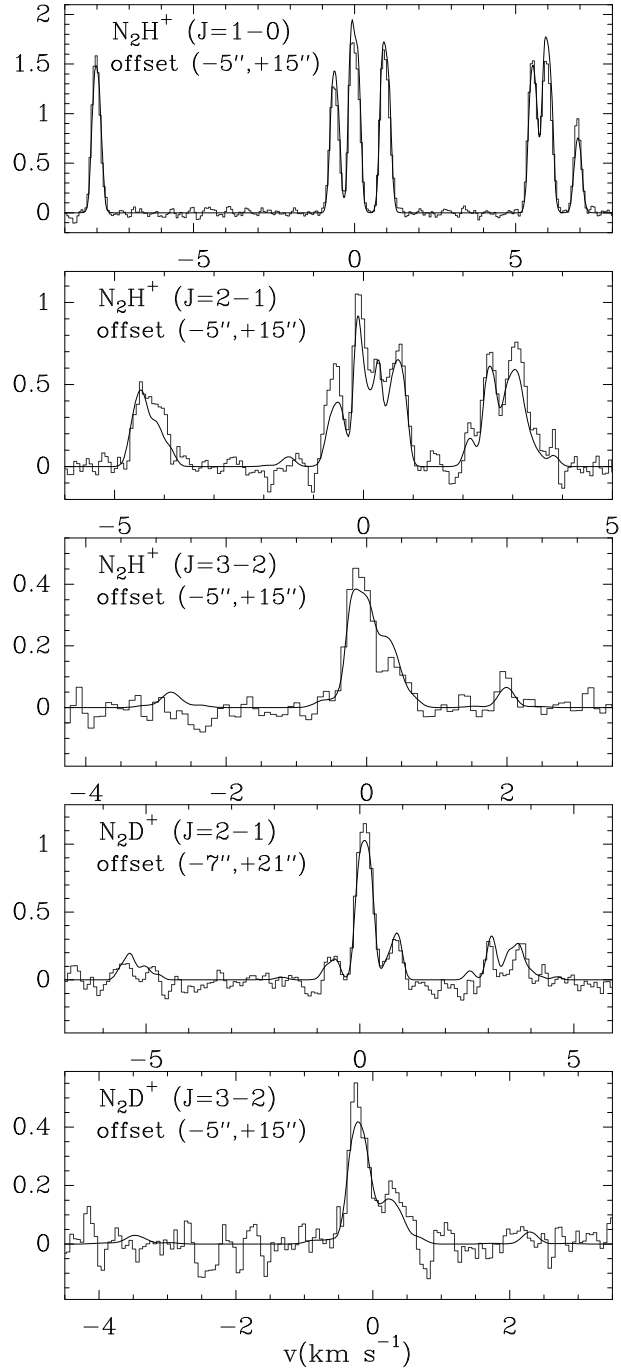}
\caption{ 
Observed and modeled $J$=1--0, 2--1 and 3--2 lines of N$_2$H$^+$ and
$J$=2--1 and 3--2 lines of N$_2$D$^+$ towards molecular peak C in L183.\label{graph:L183}}
\end{figure}

\begin{figure}
\centering
\includegraphics[scale=0.7,angle=0]{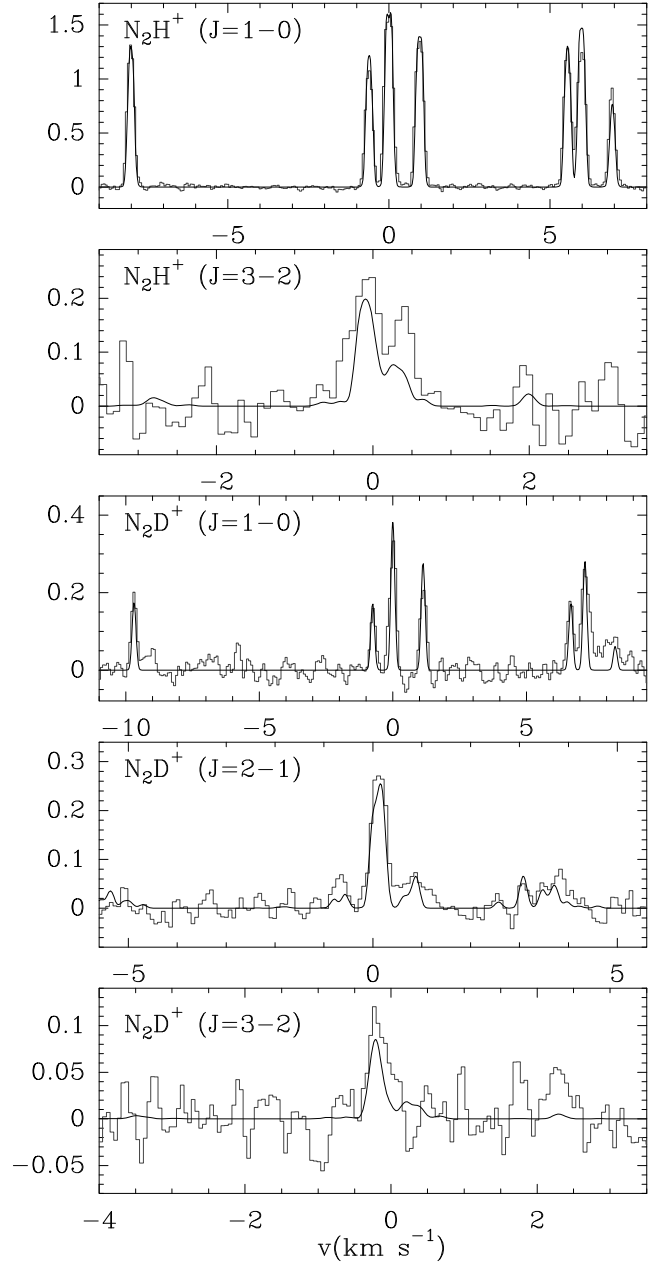}
\caption{ 
Observed and modeled $J$=1--0 and 3--2 lines of N$_2$H$^+$ and 
$J$=1--0, 2--1 and 3--2 lines of N$_2$D$^+$ for the molecular peak N in L183.
\label{graph:L134N}}
\end{figure}

\begin{figure}
\centering
\includegraphics[scale=0.7,angle=270]{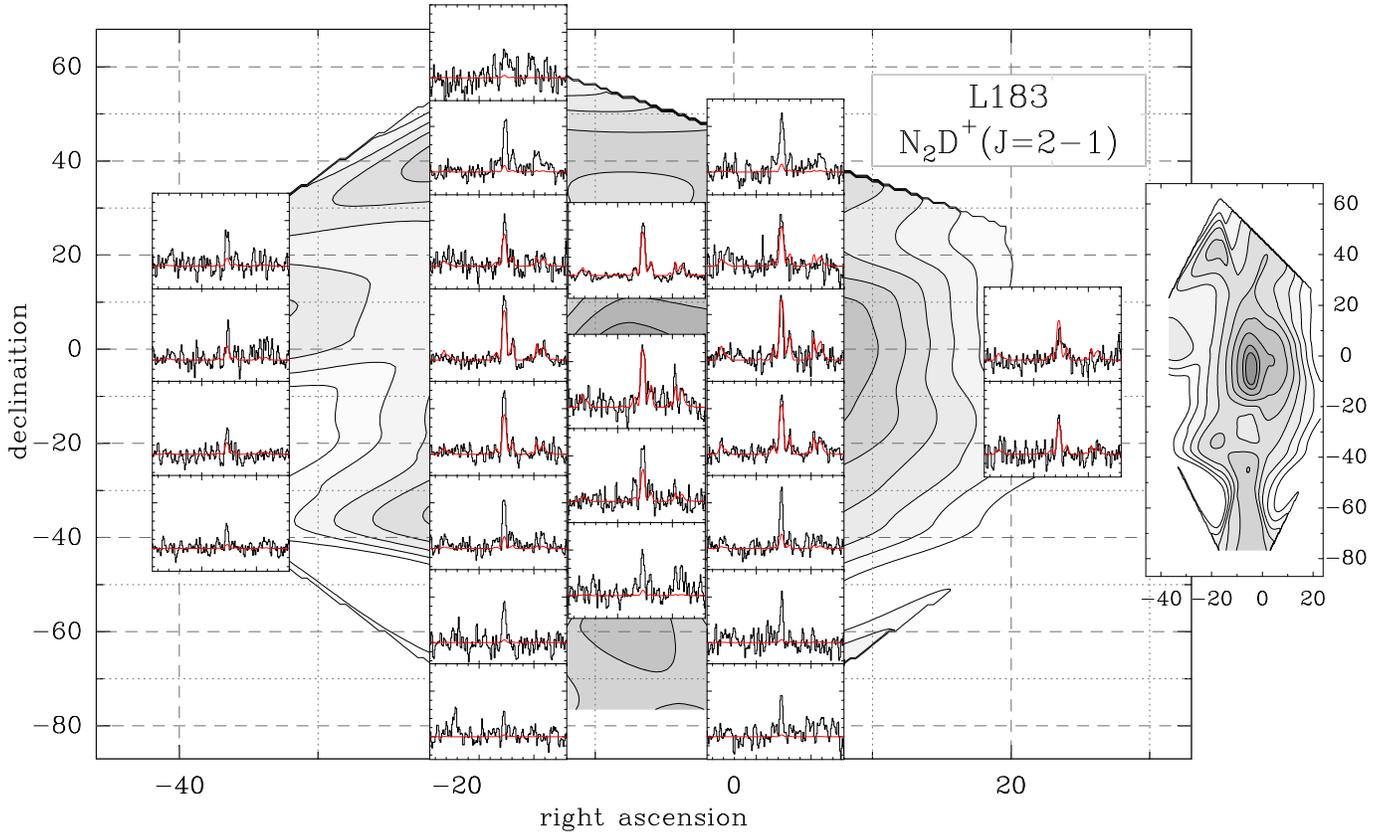}
\caption{ 
Map and modeled lines of the $J$=2--1 transition of N$_2$D$^+$ corresponding to 
molecular peak C in L183. The reference position of the map is reported in 
table \ref{table:obs_mod} and corresponds to the N$_2$H$^+$ and N$_2$D$^+$ observations shown on
figure \ref{graph:L183}. On the right side of the main map, the map of integrated intensity
is reported with contour levels going from 20\% to 90\% of the map peak intensity, by step of 10\%.
Note that the aspect ratio between right ascension and declination is respected on the right map and 
show the South--North extent of the cloud.
\label{graph:map_L183_D21}}
\end{figure}

\begin{figure}
\centering
\includegraphics[scale=0.7,angle=270]{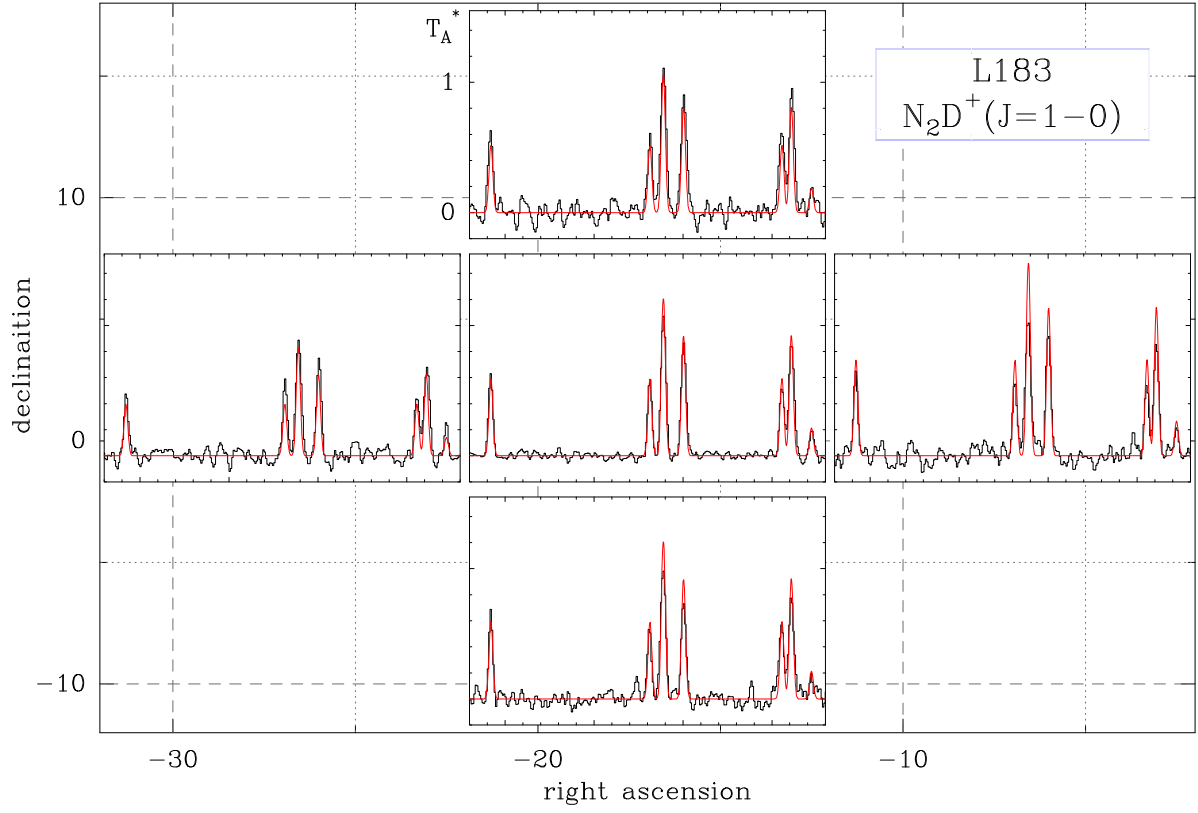}
\caption{ 
Map and modeled lines of the $J$=1--0 transition of N$_2$D$^+$ corresponding to 
the molecular peak C in L183. 
\label{graph:map_L183_D10}}
\end{figure}

Figure \ref{graph:L183} shows peak C observations and models of the $J$=1--0, 2--1 and 
3--2 transitions of N$_2$H$^+$, and of the $J$=3--2 transition of N$_2$D$^+$. 
Figure \ref{graph:L134N} 
shows the $J$=1--0 and 3--2 lines of N$_2$H$^+$ and the $J$=1--0, 2--1 and 3--2 lines of 
N$_2$D$^+$ for peak N. Figures \ref{graph:map_L183_D21}  
and \ref{graph:map_L183_D10} present maps of the $J$=1--0 and 2--1 lines of 
N$_2$D$^+$ obtained towards peak C, the latter being obtained with the 
same data as in \cite{tin00}.

There are a number of studies regarding the dust emission properties of L183 (e.g.
\citet{kir05} at 450 and 850 $\mu$m, \citet{juv02}, \citet{pag03} at 200 $\mu$m and 
\citet{pag05} at 1.2 mm). From these studies, it seems that the region of molecular
peak C is associated with two continuum sources FIR1 (south) and FIR2 (north) separated by $\sim$ 1.5'. 
The region of FIR1 and FIR2 corresponds to the maximum extinction and minimum  
temperature in the cloud \citep{juv02}. 
Also, there might be a gradient of temperature in this region or changes in the 
composition of the grains since FIR1 is detected at 200 and 450 $\mu$m but remains 
undetected at 850 $\mu$m. Moreover, the emission peak FIR2 
at 850 $\mu$m is shifted to the North 
compared to that at 200 and 450 $\mu$m \citep{kir05}, while the 1.2 mm peak position
is shifted to the South. It is worth noting that the maxima of the 1.2 mm 
and N$_2$H$^+$ maps reported in \citet{pag05} do not coincide. From this study, it appears 
that the maxima of these two maps are shifted by $\sim$ 25'' and that the H$_2$ volume 
density reaches 2 10$^6$ cm$^{-3}$ at the position of the 1.2 mm map emission peak. 
For simplicity, we have modeled the observations of peak C assuming that the density center
of our model is at the position of the N$_2$H$^+$ emission peak (i.e. offset by (+5",-15") with 
respect to the coordinates given in table \ref{table:obs_mod}).

For the spectra belonging to peak N, our observations are off by (+2'',-40'') 
with respect to the reference position given by \citet{dic00}. We have assumed in our models
that the density peak is centered at the position of our observations which
corresponds to the secondary maximum in the N$_2$H$^+$ map.

For peak C, in order to fit the observed asymmetry in the $J$=2--1 transition,
we have included a velocity gradient of $\Delta$v = -0.2 km s$^{-1}$ pc$^{-1}$. 
As discussed in \citetalias{dan06}, this line 
is sensitive to small velocity gradients, due to the presence of multiple hyperfine components
in a small frequency interval. For peak N, no velocity 
gradient was included since, for this position, the observed lines are insensitive to 
such small velocity fields (the $J$=2--1 line was not observed).
It is worth noting that the $J$=2--1 transition of N$_2$H$^+$ obtained towards 
peak C has a peculiar shape that is not observed 
in any other cloud of the present sample. Hence, the abundance profile 
obtained for this source is unique, among all our models, since we find that 
the best agreement between models and observations is obtained by increasing the abundance 
at large radii. 

For the two peaks, we obtain similar  N$_2$H$^+$ abundances, but different
density profiles: peak C is more dense and centrally peaked than peak N. 
For the former position, our density estimate is consistent with 
that obtained from sub--mm observations, i.e. n(H$_2$) = 10$^6$ cm$^{-3}$ \citep{kir05}.

In order to model the N$_2$D$^+$ map shown in Figures \ref{graph:map_L183_D21}
and \ref{graph:map_L183_D10}, the relative abundance has been set to $r_D$ = 0.43 for radii 
below r $\sim$ 0.021 pc. For larger radii, the relative abundance is lowered to 0.05. 
Note that this decrease in abundance is necessary to reproduce the line shapes as it 
reduces the self--absorption induced by the external layers of the cloud, in the $J$=2--1 
main component of N$_2$D$^+$.
On the contrary, the main component of the $J$=2--1 transition of N$_2$H$^+$ is strongly 
self--absorbed and is better reproduced by increasing the N$_2$H$^+$ abundance 
at the outermost radii. Since our model considers spherical geometry while the 
morphology of the cloud is extended along a 
North--South axis, we have failed to reproduce the $J$=2--1 transition for the whole map.
Hence, the intensities 
are well reproduced for declinations between $\delta$ = -20'' and +20'' but are underestimated 
along the North--South axis. 
Toward the position given in table \ref{table:obs_mod},
we find a column density ratio
R = 0.33$^{+0.16}_{-0.11}$ which is similar to the one obtained by \citet{tin00} at a position offset by $\sim$ 5" to the north, i.e. R = 0.35.
This value is higher than the one obtained by \citet{cra05} at a position offset by (-5",-15"), i.e. R = 0.22 $\pm 0.04$. 

From the observations towards peak N, we obtain values for the deuterium
enrichment of $r_D$ = 0.76 for radii below 5 10$^{-3}$ pc and 0.05 beyond. Note that
the radius which encloses the central D--rich region is determined 
from a single position analysis. Because the angular size of this central region 
is intermediate between the HPBW of the $J$=1--0 and 3--2 lines of N$_2$D$^+$, 
(32'' and 10'' respectively), the $J$=3--2 line is favored compared to the 
$J$=1--0 one. In order 
to better constrain the variation of the D/H ratio with radius, observations at 
different positions are needed. For peak N, we derive a column density ratio of 
R = 0.15 .
 
Recently, a similar work based on N$_2$H$^+$ and N$_2$D$^+$ observations was carried out 
by \cite{pag07} for molecular peak C. This latter study differs from ours in the location of the center 
of the density profile: it was set coincident with the maximum of the 1.2 mm map which is shifted 
$\sim$ 25" to the south compared to that assumed in the present work.
Moreover, the 1D radiative transfer code used in \citet{pag07} work treats line overlap and the 
convolution with the telescope beam was made using an approximation that takes into account the north-south 
extent of the cloud. At the position of the 1.2 mm map peak, they found that the 
temperature ranges from 7K (from N$_2$H$^+$ data) to 8K (from N$_2$D$^+$ data) and that
the maximum H$_2$ density is 3 times higher than that derived in the present work.
Note that the density profile derived here and the "best model" profile in \cite{pag07} 
are similar beyond 30": the latter corresponds to densities which are higher by $\sim$ 25\%. 
The main difference concerns the N$_2$H$^+$ abundance profile. \cite{pag07} have found that X(N$_2$H$^+$) 
drops by a factor $6^{+13}_{-3}$ in the inner 20" region. This decrease is interpreted as depletion and 
it becomes efficient at densites
n(H$_2$) = 5--7 10$^{5}$ cm$^{-3}$. At our position, we do not see evidence for such a strong decrease in abundance:
the abundance we find is $\sim$ 1.6 time lower below 40" where the density reach 
n(H$_2$) = 2 10$^{5}$ cm$^{-3}$. Given that the source strongly departs from spherical geometry, this variation 
could be a consequence of a mis--interpretation of the real excitation conditions and prevents us to interpret 
this result as a sign of depletion. Considering the deuterium enrichment, \cite{pag07} find that at the 1.2 mm map peak, 
$r_D$ varies from 0.7 $\pm 0.12$ (inner 6") to 0.05 (beyond 40"), which is consistent with the present findings.

\subsubsection{TMC2}
\begin{figure}
\centering
\includegraphics[scale=0.7,angle=0]{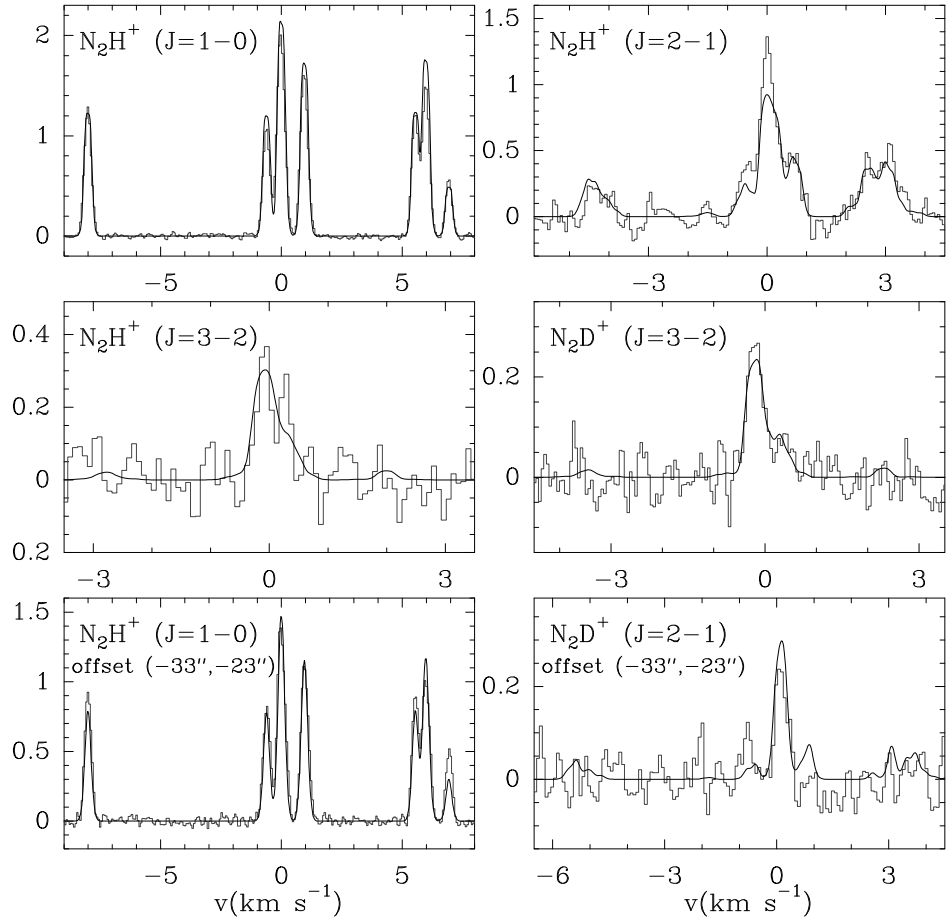}
\caption{ 
Observed and modeled lines for two positions toward TMC2.
At the position of the N$_2$H$^+$ map peak are shown the 
$J$=1--0, 2--1 and 3--2 lines of N$_2$H$^+$ and the 
$J$=3--2 line of N$_2$D$^+$. For the offset position, are shown
the $J$=1--0 line of N$_2$H$^+$ and the $J$=2--1 line of N$_2$D$^+$.
\label{graph:TMC2}}
\end{figure}

In the present study, we report N$_2$H$^+$ and N$_2$D$^+$ observations for two 
positions towards TMC2 which are separated by $\sim$ 40''. Figure \ref{graph:TMC2} 
shows the $J$=1--0, 2--1 and 3--2 observed lines of N$_2$H$^+$, and the $J$=3--2 line of N$_2$D$^+$ 
for the central position. For the (-33'',-23'') position, 
N$_2$H$^+$ $J$=1--0 and N$_2$D$^+$ $J$=2--1 lines are shown. Figure \ref{graph:int_rad}
shows the $J$=1--0 radial integrated intensity of the model compared
to observational data from \citet{cas02}. 

This source has been mapped in many molecular lines (e.g. C$^{18}$O, DCO$^+$ 
and H$^{13}$CO$^+$ by \citet{but95}, H$^{13}$CO$^+$ by \citet{oni02} and 
C$^{18}$O by \citet{oni96}) as well as in 1.2 mm continuum emission by \citet{cra05}. 
Large scale visual extinction maps covering the whole region are reported by \citet{cer84,cer85b}. 
Contrary to other clouds, N$_2$H$^+$ and 1.2 mm maps seem to be uncorrelated: while the 
N$_2$H$^+$ peak has a bright 1.2 mm emission counterpart, there are several maxima in the 
1.2 mm map which do not correspond to any N$_2$H$^+$ peak. 
In this work, we derive an average central volume density of n(H$_2$) $\sim$ 3.5 10$^5$ cm$^{-3}$,
which is in good agreement with the value of 3 10$^5$ cm$^{-3}$ derived from the
analysis of the 1.2 mm emission. However, this value is higher than the estimates obtained from 
other molecular species. The discrepancy partly arises from a different spatial location
of the molecules (see the map of C$^{18}$O reported by \citet{cra05} and the map of 
H$^{13}$CO$^+$ of \citet{oni02}). On the other hand, N$_2$H$^+$ and NH$_3$ lines \citep{mye79} 
seem to arise from the same region and an average volume density of 2.5 10$^4$ cm$^{-3}$ was derived 
from the latter molecule. Similar values for the average density are given in \citet{but95}, based 
on C$^{18}$O and DCO$^+$ lines. Nevertheless, higher density estimates have been
obtained using H$^{13}$CO$^+$ \citep{oni02}. The map of this molecule shows that 
two emission maxima are located within the NH$_3$ map corresponding to regions 
of average volume density $\sim$ 10$^5$ cm$^{-3}$. 

From NH$_3$ lines, the mass enclosed within a radius of 3.6' has been
estimated to be $\sim$ 16 M$_{\sun}$, and from C$^{18}$O lines \citep{oni02}, the mass 
enclosed within 7.8' is estimated to be $\sim$ 50 M$_{\sun}$.
The density profile given in table \ref{table:models} corresponds to enclosed 
masses of 50 M$_{\sun}$ and 121 M$_{\sun}$ for the same radii. We stress that the 
value adopted for r$_0$ is the most important parameter
to derive the mass of the cloud and, in order to obtain similar values than those quoted above
and using the same central density, we would have to adopt a value of r$_0$ $\sim$ 35''. Adopting 
such a small value for r$_0$ does not allow to reproduce the $J$=1--0 radial integrated 
intensity. 

To reproduce the N$_2$D$^+$ spectra obtained towards the two positions, we have 
decreased the abundance ratio by a factor 7 at radii $>$ 0.024 pc. 
It is worth noting that a single position analysis enables to 
reproduce independently each one of the two N$_2$D$^+$ spectra, assuming a 
constant value for $r_D$. 
For the central position, such an analysis leads to a ratio of $r_D$ $\sim$ 0.25. 

\subsubsection{TMC1(NH$_3$)}

\begin{figure}
\centering
\includegraphics[scale=0.7,angle=0]{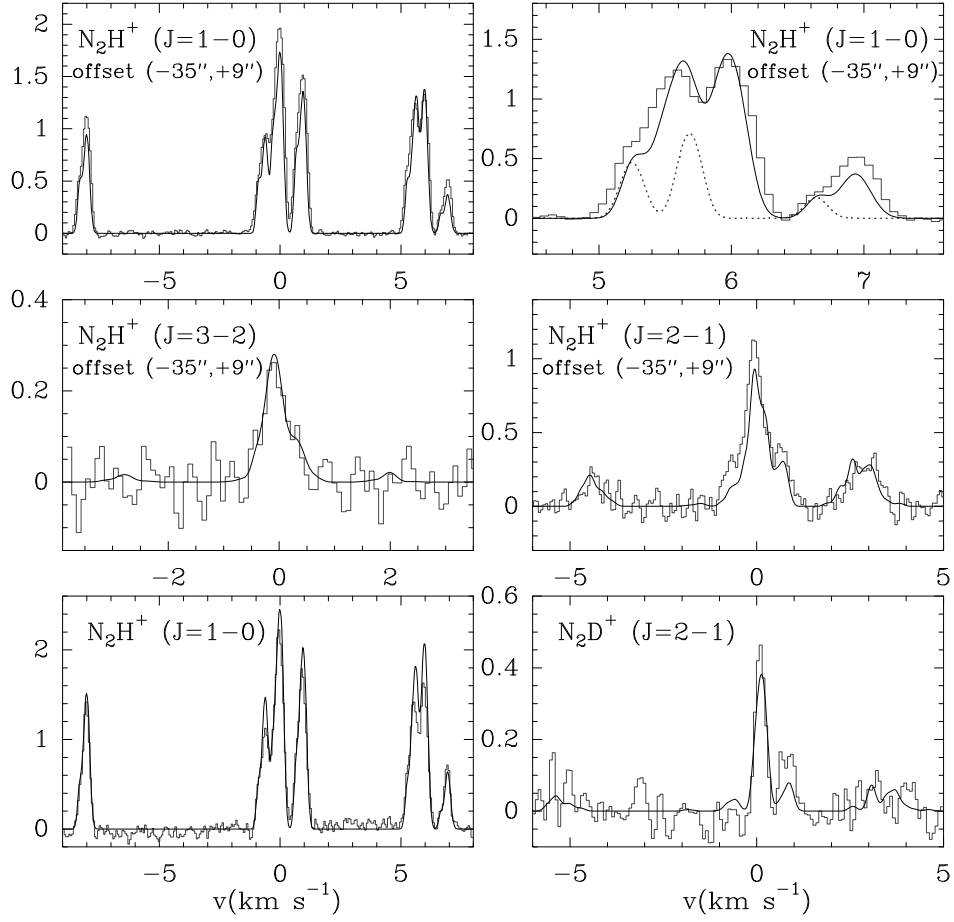}
\caption{ 
Models for the $J$=1--0, 2--1 and 3--2 line of N$_2$H$^+$ and for the
$J$=3--2 line of N$_2$D$^+$ observed in TMC1(NH$_3$). The upper right
panel is a blow--up of the low frequency hyperfine triplet in the $J$=1--0
line, with the second emission component indicated.
\label{graph:TMC1-NH3}}
\end{figure}

Figure \ref{graph:TMC1-NH3} shows observations of N$_2$H$^+$ and 
N$_2$D$^+$ towards two positions separated by $\sim$ 35''.
The reference position is given in table \ref{table:obs_mod} and the second
position is located at (-35'',+9''). Towards both positions, two velocity components
are identified in the observations of the $J$=1--0 line of N$_2$H$^+$. 
In order to model the faintest component, we have simply added the contribution of
two emitting regions. This way of computing the emerging spectra is motivated 
by the fact that the difference in the V$_{LSR}$ of the two components
is larger than the linewidth: the determined V$_{LSR}$ 
are 5.98 km $s^{-1}$ and 5.68 km $s^{-1}$, the former being the most intense one. 
For both positions, the faintest velocity components have been determined to be similar.

The TMC1 ridge is known to have a complex velocity structure with a velocity gradient
of $\sim$ 0.2 km s$^{-1}$, perpendicular to the major axis of the ridge \citep{ola88} ,
and cores at different V$_{LSR}$ for the same line of sight. 
Particularly, two components at $\sim$ 5.6 km $s^{-1}$ 
and 6.0 km $s^{-1}$ have been identified around the cyanopolyyne peak \citep{tol81}. 
Also, around the ammonia peak, two possible components 
at V$_{LSR}$ $\sim$ 5.9 km $s^{-1}$ and 5.3--5.4 km $s^{-1}$
are reported by \citet{hir92} from observations of C$^{34}$S transitions, 
the latter being identified as a low density core. The two V$_{LSR}$ components 
identified in this work are also seen in SO observations \citep{liq06}.

As for the previous sources, we obtain a higher density estimate than in previous
studies based on the emission from other molecules. See for example the detailed analysis of the 
density structure along the TMC1 ridge reported by \citet{pra97}. At the position of 
the N$_2$H$^+$ peak (ammonia reference position), the density is 
estimated to be $\sim$ 10$^5$ cm$^{-3}$ from the analysis of several HC$_3$N lines. Note that 
the structure of the cyanopolyyne peak has been probed through interferometric 
observations \citep{lan95}. It has been found that the core of the cloud consists 
of several high density condensations (n(H$_2$) $\sim$ 3 10$^5$ cm$^{-3}$) with diameters 
ranging from 10'' to 30'', which are embedded in a lower density medium. 
This result was further confirmed by \citet{tot04} who analyzed ISOPHOT images.  

From the $J$=2--1 line of N$_2$D$^+$, we derive R $\sim$ 0.10$^{+0.09}_{-0.04}$ for the 
ammonia reference position, which is in good agreement with the value 
of 0.08 reported by \citet{tin00}.

\subsubsection{L1498}
\begin{figure}
\centering
\includegraphics[scale=0.7,angle=0]{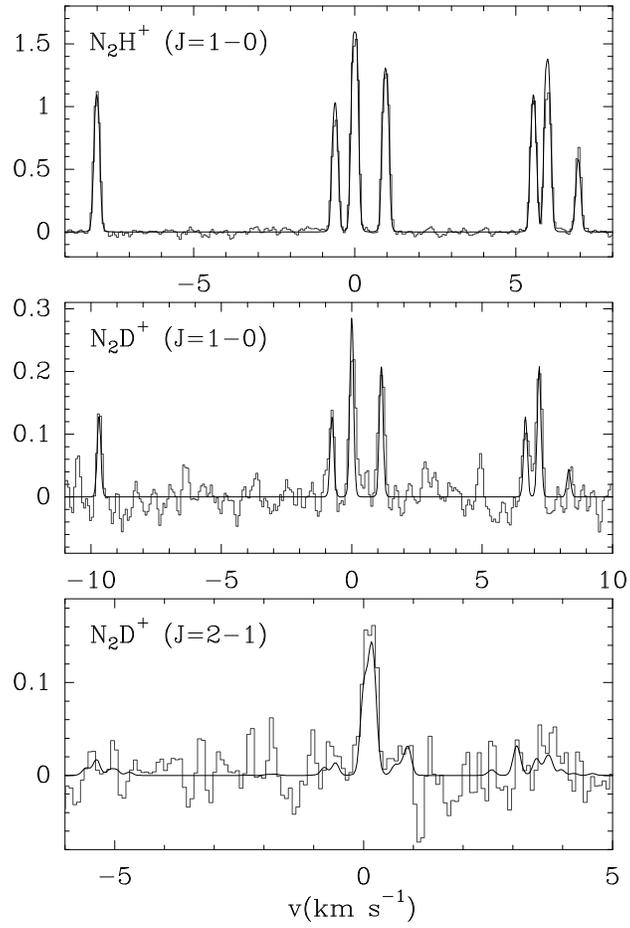}
\caption{ 
Observed and modeled $J$=1--0 line of N$_2$H$^+$, and the $J$=1--0 and 2--1 lines of 
N$_2$D$^+$ towards L1498.
\label{graph:L1498}}
\end{figure}

Figure \ref{graph:L1498} shows the $J$=1--0 line of N$_2$H$^+$, and the $J$=1--0 and 2--1 
lines of N$_2$D$^+$. 
L1498 is a nearby molecular cloud located in the Taurus complex.
This cloud has been observed in H$_2$CO \citep{you04}, CS \citep{taf04}
and N$_2$H$^+$ \citep{taf04,shi05} and its physical 
properties have been derived from ISOPHOT observations \citep{lan01} and 
millimeter observations \citep{shi05}.

As for L1517B, a detailed study of L1498 has been done by \citet{taf04},
and we have used this work as a starting point for our models. The 
position we observed in this source is at (+10'',0'') of the 1.2 mm emission peak. 
In our model, we have assumed that the center of the density profile 
corresponds to the continuum peak.
Figure \ref{graph:int_rad} shows the $J$=1--0 radial integrated
intensity compared to the observational data reported by \citet{taf04}. 
As noted in that study, a decrease of the abundance in the outer envelope 
is needed to reproduce the $J$=1--0 radial intensity profile. 
Otherwise, the intensity would be overestimated at large radii. 
 
The values we have derived for the density and for the N$_2$H$^+$ abundance are consistent 
with those found by \citet{taf04}. As for L1517B, we derive a 
higher abundance than in the latter study (X(N$_2$H$^+$)= 2.5 10$^{-10}$ 
compared to 1.7 10$^{-10}$). This is related to the use of different collisional 
rate coefficients and density profiles in each study.
Moreover, we derive the same central density, but with a smaller value for 
r$_0$ due to the different analytical expressions of the density profile. Nevertheless, the central density reported in this work 
is higher than what is usually found for this cloud. For example, \citet{shi05} reported 
a density $\sim$ 1--3 10$^4$ cm$^{-3}$ from continuum observations at 350 $\mu$m, 850 $\mu$m 
and 1.2 mm.  

The value we have found for the column density ratio, i.e. R = 0.07$^{+0.03}_{-0.03}$,
 is larger by a factor
two than the one reported by \citet{cra05} at the same position.

\subsubsection{L63}
\begin{figure}
\centering
\includegraphics[scale=0.7,angle=0]{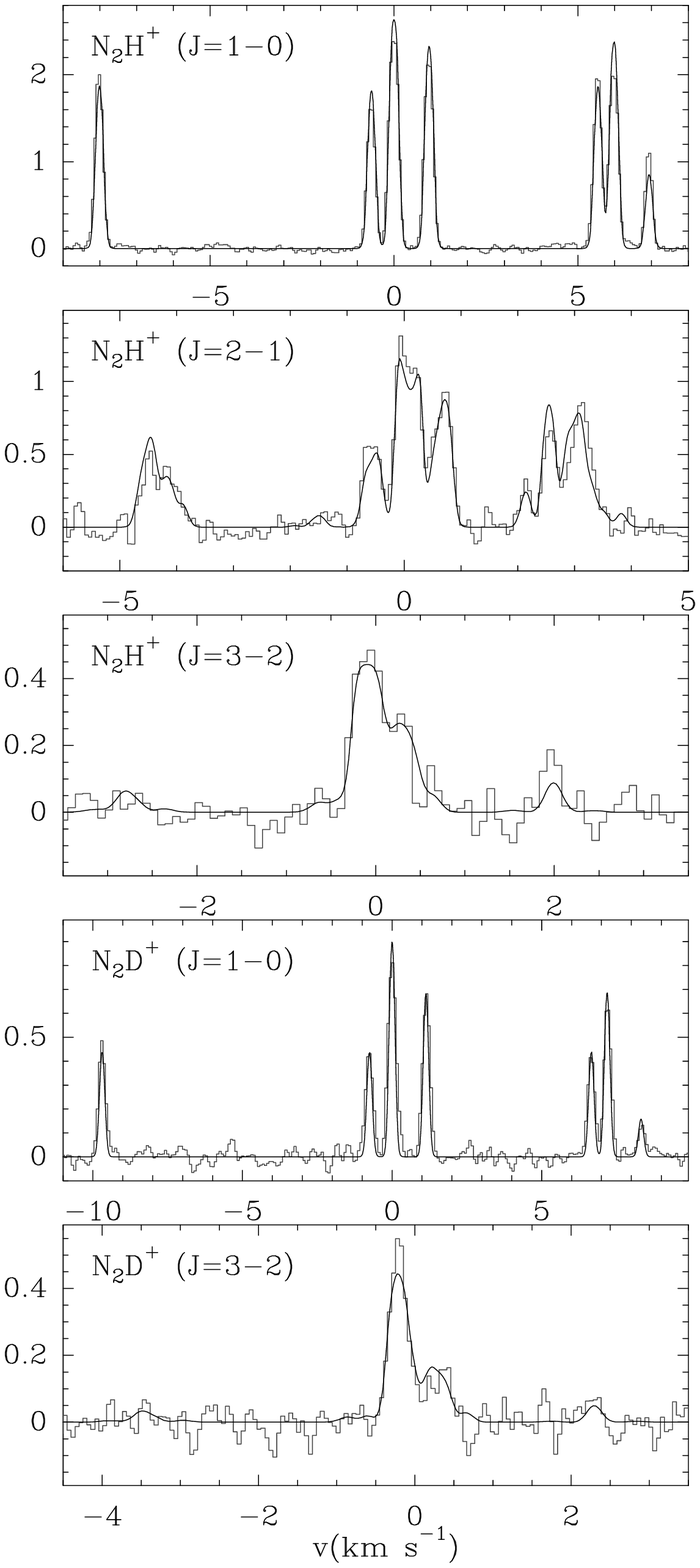}
\caption{ 
Observed and modeled $J$=1--0, 2--1, and 3--2 lines of N$_2$H$^+$ and 
$J$=1--0 and 3--2 lines of N$_2$D$^+$ towards L63.
\label{graph:L63}}
\end{figure}

\begin{figure}
\centering
\includegraphics[scale=0.7,angle=270]{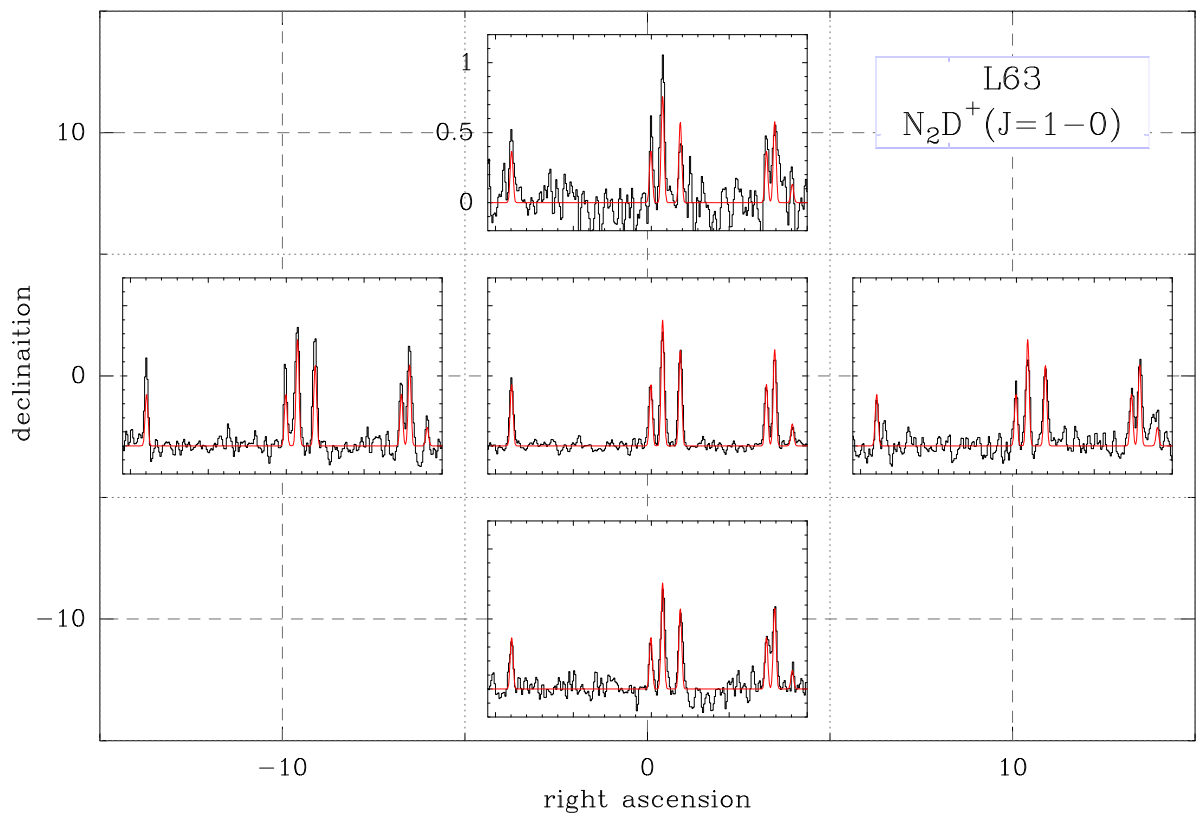}
\caption{ 
Observed and modeled J=1--0 lines of N$_2$D$^+$ for five position towards L63.
\label{graph:map_L63}}
\end{figure}

Figure \ref{graph:L63} presents the $J$=1--0, 2--1 and 3--2 lines of N$_2$H$^+$ and the 
$J$=1--0 and 3--2 lines of N$_2$D$^+$. 
Figure \ref{graph:map_L63} shows a 5 points map of the $J$=1--0 transition of N$_2$D$^+$ 
(the reference position is given in table \ref{table:obs_mod}).
Figure \ref{graph:int_rad} compares the $J$=1--0 radial integrated 
intensity of the model to observational data reported by \citet{cas02}.

In order to simultaneously reproduce the $J$=1--0 and 3--2 lines of N$_2$D$^+$, 
a step in the isotopologues abundance ratio has been introduced, 
this ratio varying from $\sim$ 0.6$^{+0.5}_{-0.3}$ in the inner 10'' to 
$\sim$ 0.25$^{+0.16}_{-0.02}$ at greater radii. Note that the extent of the enriched region favors the intensity
in the $J$=3--2 line compared to the $J$=1--0 line (see discussion of L183).

L63 has been mapped in NH$_3$ \citep{ben89} , N$_2$H$^+$ and CO \citep{sne81} 
as well as at 800 $\mu$m and 1.3 mm continuum emission \citep{war94,war99}. 
The intensity peak of both the continuum emission and N$_2$H$^+$ maps are found 
at the same position. The central density and break radius obtained in this work 
(n$_0$ = 7.2 10$^5$ cm$^{-3}$, r$_0$ = 0.018 pc) are in excellent agreement with the 
corresponding values derived by \citet{war99} from the analysis of 1.3 mm emission
(n$_0$ = 7.9 10$^5$ cm$^{-3}$, r$_0$ = 0.017 pc). In comparison to the results derived 
by \citet{ben89} from NH$_3$ observations, the profile determined in this work corresponds 
to a more massive cloud (within 2.4', 8 M$_{\sun}$ from NH$_3$ compared to 
16.5 M$_{\sun}$ from N$_2$H$^+$). Such a discrepancy seems to be present all sources 
we have modeled. 

For this source, we have introduced a radial variation of the temperature, 
going from 8 K in the inner part to 15 K in the outer envelope. 
This is consistent with previous works where the temperature has been estimated 
to be 8 K in the inner 40'' (continuum, \citet{war94}), 
9.7 K inside 2.4' (NH$_3$, \citet{ben89}), and 15 K within 8' (CO, \citet{sne81}).
Such an increase of the temperature in the outer regions of clouds 
is expected due to heating of the gas and dust by the interstellar radiation
field.

\subsubsection{L43}

\begin{figure}
\centering
\includegraphics[scale=0.7,angle=0]{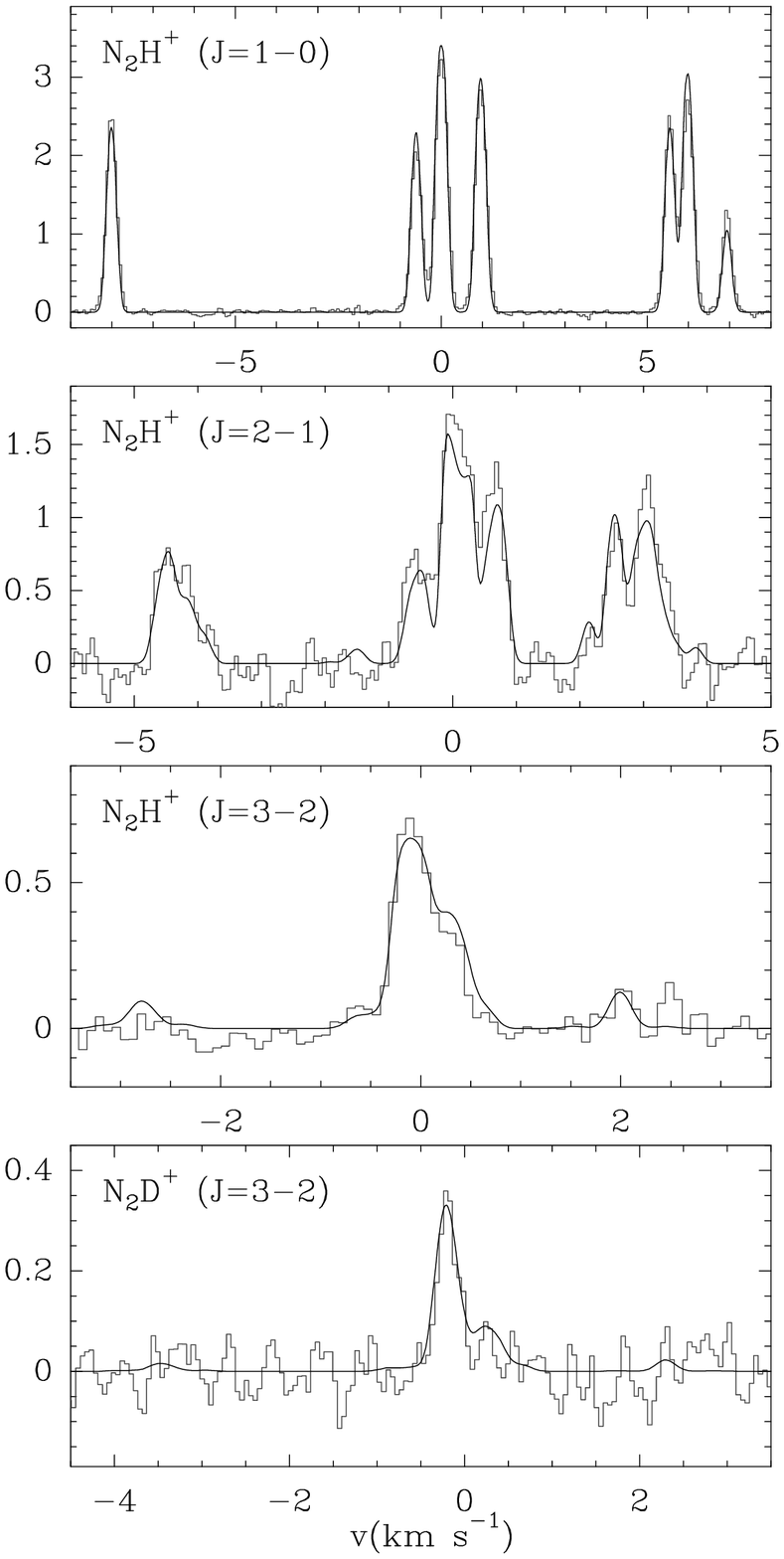}
\caption{ 
Observed and modeled $J$=1--0, 2--1, and 3--2 lines of N$_2$H$^+$ and 
$J$=3--2 line of N$_2$D$^+$ towards L43.
\label{graph:L43}}
\end{figure}

Figure \ref{graph:L43} shows the $J$=1--0, 2--1 and 3--2 lines of N$_2$H$^+$ and the 
$J$=3--2 line of N$_2$D$^+$, and Figure \ref{graph:int_rad} presents the $J$=1--0 radial integrated 
intensity from the model compared to the observations reported by \citet{cas02}. 

The dense core of L43 is composed of one main condensation with a filamentary extent
at the South--East. This core is associated with the T--tauri star RNO91 located 1.5' West. 
As pointed out by \citet{war99}, the initial cloud might be forming multiple 
stars, RNO91 being the first to have appeared.  From 1.3 mm continuum emission, 
\citet{war99} found an average central volume density of 
$\sim$ 2 10$^6$ cm$^{-3}$ within a radius of r$_0$=0.018 pc, at which a steeper
slope is reached. The density profile we have determined is less centrally peaked 
(i.e r$_0$ = 0.022 pc, n$_0$ = 9.2 10$^5$ cm$^{-3}$) but leads to the same estimate of 
the enclosed mass in the cloud: we obtain masses of 1.1 and 8.2 M$_{\sun}$ within, respectively, 22'' 
and 51''. This latter radius corresponds to the geometric average
of the FWHM major and minor axis quoted by \citet{war99} at a distance of 160 pc. 
Our results are in excellent agreement with the estimations of 
1.2 and 8.2 M$_{\sun}$ of the latter work.

\subsubsection{L1489}
\begin{figure}
\centering
\includegraphics[scale=0.7,angle=0]{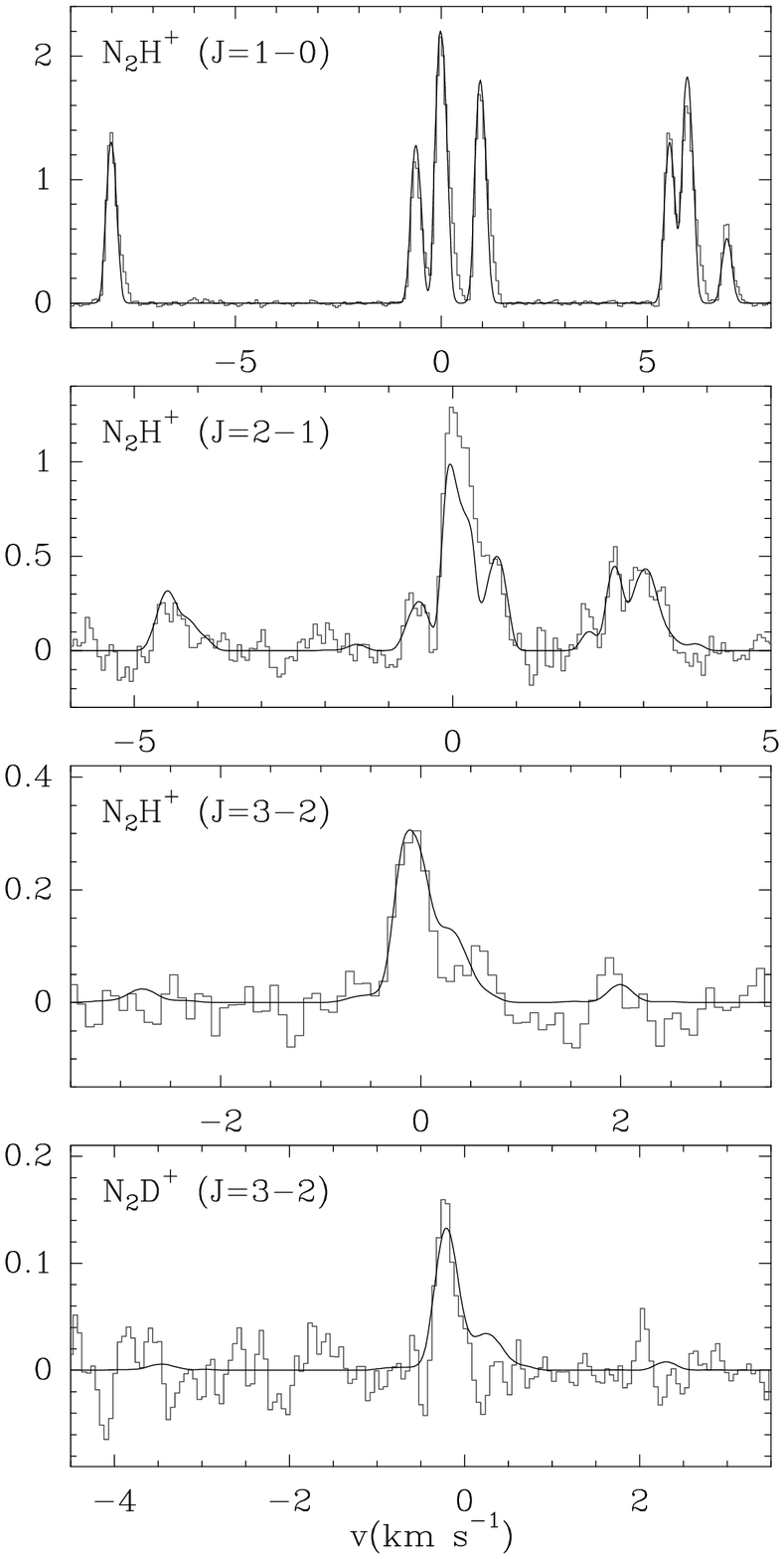}
\caption{ 
Observed and modeled $J$=1--0, 2--1 and 3--2 lines of N$_2$H$^+$ and
$J$=3--2 line of N$_2$D$^+$ towards L1489.
\label{graph:L1489}}
\end{figure}

Figure \ref{graph:L1489} presents the $J$=1--0, 2--1 and 3--2 lines of N$_2$H$^+$ and the 
$J$=3--2 line of N$_2$D$^+$. Figure \ref{graph:int_rad} shows the $J$=1--0 radial integrated 
intensity compared to observational data obtained by \citet{cas02}. 
Our observation of the $J$=1--0 line shows asymmetrical line wings with an enhanced 
red wing that we failed to reproduce in our models. Thus, the integrated intensity shown 
on Figure \ref{graph:int_rad} is underestimated in the model by comparison 
to observational data.

This source belongs to the Taurus complex and is associated with the low mass YSO, IRAS 04016+2610 
situated $\sim$ 1.2' to the West. The NH$_3$ map shows one main condensation of 
radius $\sim$ 0.07 pc with a peak intensity shifted $\sim$ 1' North of the N$_2$H$^+$ 
peak. \citet{ben89} estimated the temperature to be 9.5--10 K and 
the main condensation mass to be around 1.6--2.1 M$_{\sun}$, which is lower than our estimate of 6.5 M$_{\sun}$. 
On the other hand, observations of HCO$^+$ lines performed by \citet{oni02} 
have shown that two condensations are lying in the NH$_3$ emitting region. 
These clumps have a low spatial extent (r = 0.018 and 0.016 pc ) and 
correspond to dense cores with masses $\sim$ 1 M$_\sun$. 

\subsubsection{L1251C}

\begin{figure}
\centering
\includegraphics[scale=0.7,angle=0]{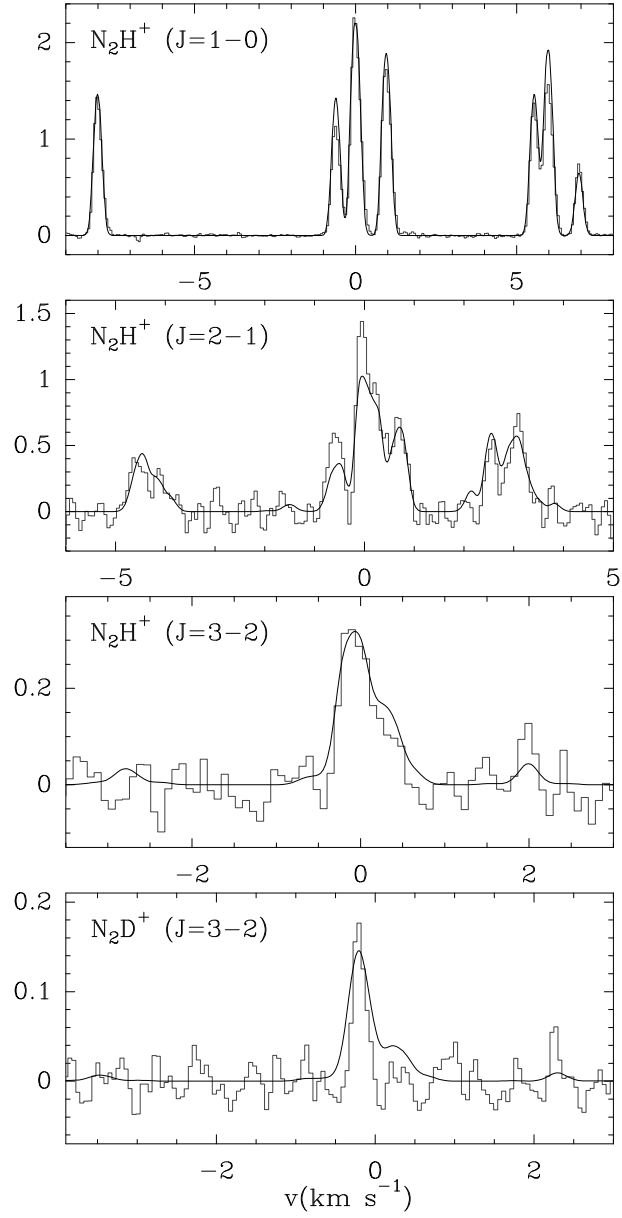}
\caption{ 
Observed and modeled $J$=1--0, 2--1 and 3--2 lines of N$_2$H$^+$ and 
$J$=3--2 line of N$_2$D$^+$ towards L1251C.
\label{graph:L1251C}}
\end{figure}

Figure \ref{graph:L1251C} shows the $J$=1--0, 2--1 and 3--2 lines of N$_2$H$^+$ and the 
$J$=3--2 line of N$_2$D$^+$. 

L1251C is a dense core belonging to a region of active stellar formation in the Cepheus complex.
Large scale studies of this region are reported for $^{13}$CO and C$^{18}$O by \citet{sat94} and for 
NH$_3$ by \citet{tot96}. 
As for the other clouds, we have
determined a higher density from our N$_2$H$^+$ data compared to that obtained from analysis
based on other molecular species. This results in a larger mass estimate for the cloud. In previous studies,
masses of 5 M$_\sun$ within 62'' \citep{tot96} and 56 M$_\sun$ within 
300'' \citep{sat94} have been obtained. In this work, we derive masses of 13 
and 87 M$_\sun$ respectively.

\subsection{Discussion}

For all sources where the $J$=1--0 line of N$_2$D$^+$ was observed,
we have been able to obtain models which reproduce the observed hyperfine 
intensity ratios. On the other hand, it is often more difficult
to obtain such a good agreement when trying to model the $J$=1--0 
hyperfine line intensities of N$_2$H$^+$. For this molecule,
a number of studies have reported intensity anomalies for
the $J$=1--0 hyperfine transitions \citep[see e.g.][]{cas95}.
Between the two isotopologues, the main difference is due to optical 
thickness. Indeed, when comparing the observations of the $J$=2--1 lines of 
N$_2$H$^+$ and N$_2$D$^+$, we see that, for most sources, we observe 
self absorption features in the $J$=2--1 main central component of 
N$_2$H$^+$ while not for N$_2$D$^+$. Thus, this fact supports the idea that 
hyperfine intensity anomalies are due to radiative effects such as 
scattering in the low density regions of the clouds. Hence, the lower
opacity of the $J$=1--0 line of N$_2$D$^+$ compared to N$_2$H$^+$ explains 
why such intensity anomalies are not observed for the deuterated isotopologue.

For most sources, we find a better agreement between models and observations
using a N$_2$H$^+$ abundance which varies with radius. The N$_2$H$^+$ abundance 
is enhanced in the central region in all models, except for L1517B and for
molecular peak C in L183. For L183, the best agreement with observations is found 
by introducing an increase of the N$_2$H$^+$ abundance in the outer envelope. 
This could indicate a real trend for this 
source but, on the other hand, it must be confirmed with a 
more precise description of the morphology of the cloud, as previously discussed. 
Note that these two clouds are the less massive in our sample (see table \ref{table:obs_mod}). 
For all the other clouds for which we have obtained enclosed masses greater than 5 M$_{\sun}$
within 2', the N$_2$H$^+$ abundance decreases in the outermost regions. 
These different types of abundance profiles were already noticed 
for some clouds in our sample (see sect. \ref{clouds}) and this reflects the specificity 
of the chemistry in each source.

Compared to previous studies, we have derived lower N$_2$H$^+$ abundances despite 
the fact that the collisional rates used in the present work are lower than 
the rates of HCO$^+$ -- H$_2$ currently used to interpret N$_2$H$^+$
spectra. Models with high density and low abundance are the best to reproduce 
the $J$=2--1 line which suffers from self--absorption. Moreover, in the sample 
of clouds we studied, we find a weak departure in the abundances (less than a 
factor of 2) to the average N$_2$H$^+$ abundance of 1.5 10$^{-10}$. Note that for the 
same sample of clouds, the average value for the abundance reported in \citet{cas02} 
is 4 10$^{-10}$. In all these clouds, a decrease of the temperature in the 
inner region below 10 K enables to reproduce best the observations since it increases
the opacity of the $J$=1--0 line. On the other hand, an increase of the 
temperature with radius allows to minimize self absorption effects in the $J$=2--1 line. 

In our sample, several clouds have already been analyzed using continuum observations
(L63, L43, TMC--2, L183, L1498 and L1489). These studies lead to estimates 
for the mean volume density which are in good agreement with those derived 
in this work.
Nevertheless, when using other molecules, the differences 
in density estimates are more important. Part of the discrepancies may arise 
from the different spatial distribution within the clouds of each molecule. 
Some of them, such as CO or CS, are known to undergo depletion in the 
highest density regions and an analysis which assumes a uniform density 
would lead to a lower value for the cloud density. For the studies 
based on NH$_3$, the differences are yet unclear since both molecules 
seem to be present in the same regions of the clouds, as shown by the 
corresponding emission maps. 
For the comparison with the estimates done in \cite{ben89}, a source of discrepancy 
could be the larger HPBW ($\sim$ 80") of the Haystack antenna used in the latter work. 
Another possibility explaining the lower density estimates derived from NH$_3$ lines may 
come from the low critical densities of the (J,K)=(1,1) and (2,2) inversion 
lines, i.e. n$_c$ $\sim$ 10$^4$ cm$^{-3}$. Since for the densities of the inner 
regions of the dense cores, these lines are thermalized, an accurate
determination of the kinetic temperature and NH$_3$ column density is 
still possible while the lines are rather insensitive to the density.  
Moreover, the densities estimated from N$_2$H$^+$ and millimeter 
continuum observations lead to NH$_3$ line
intensities consistent with the observations \citep[see e.g.][]{taf04}. This points out
the
low sensitivity of the NH$_3$ inversion lines to dense gas, and the difficulty
to assess the dense core gas density uniquely from NH$_3$ observations.
Finally, we obtain that  
the mass derived for the clouds are in general in good agreement with the 
estimates obtained from mm or sub--mm emission analysis and higher than the mass derived 
from the observations of other molecular species.

Note, that the rate coefficients used in this work are calculated for collisions with He. 
As discussed in \citet{mon85} for the case of HCO$^+$ (isoelectronic of N$_2$H$^+$), 
the larger polarizability of H$_2$ compared to He induces strong variations in the 
corresponding potential energy surfaces. It entails that the cross sections obtained 
for HCO$^+$--H$_2$ are greater by a factor 2--3 compared to the HCO$^+$--He ones.
In the present work, the collisional rate coefficients are 
corrected for the difference in the reduced masses of H$_2$ and He in the 
Boltzmann average. This is a crude approximation which assumes that cross sections
for collisions with He and H$_2$ are similar. Thus, we expect to obtain 
significant differences, by analogy with the HCO$^+$ case, for the collisional rates
of N$_2$H$^+$ with H$_2$. 
Therefore, the results obtained have to be 
regarded with caution and the error on the density and abundance estimates could be 
as large as a factor 2. The models presented for L1517B and L1498 can serve 
as a benchmark for the typical variations that we expect using different 
collisional rate coefficients. Compared to the abundances reported in \citet{taf04}, 
the values we have derived in this work are within 50--70\%. 

From N$_2$D$^+$ lines, we have derived values for the D/H enrichment in each cloud
which are in reasonable agreement with previous determinations, despite the fact 
that the methods used are often different. 
In addition, for three of the studied clouds (L63, TMC2 and the two 
cores in L183), we find that the D/H ratio takes high values (0.5--0.7) in the inner
part of the cloud and decreases in the lower density outer regions.
Such characteristics are expected from theoretical models which couple the dynamics and 
chemistry of molecular clouds \citep{aik05,rob03}. Moreover, these models predict 
that a quantitative determination of the deuterium enrichment could probe 
the evolutionary stage of the collapse.

\section{Chemical analysis}
The present analysis of the line profiles of N$_2$H$^+$ and N$_2$D$^+$ 
has allowed to derive 
the profile and temperature dependences together with the overall deuterium 
fractionation ratio in the various selected pre--stellar cores. Looking at 
table \ref{table:models2}, one may notice that the deduced total column density of N$_2$H$^+$ 
varies within a factor of 2 whereas the deuterated counterpart abundance 
may vary by factors larger than 6, which leads to  fractionation values 
between 0.07 and  0.5.

A detailed chemical discussion on the deuterium fractionation derived from 
the present observations is beyond the scope of the present paper and we 
only want to derive some general trends from the observations.
The presence of molecular ions in interstellar environments is readily 
explained by a succession of ion molecule reactions initiated by cosmic 
ray (CR) ionization of molecular hydrogen : 
H$_2$ + CR $\rightarrow$ H$_2^+$; H$_2$ + H$_2^+$ $\rightarrow$ H$_3^+$ + H. 
The H$_3^+$ molecular ion is an 
efficient proton donor and reacts with saturated stable molecules such 
as CO, N$_2$, HD  to produce HCO$^+$, N$_2$H$^+$,   H$_2$D$^+$.  
Deuterium enhancement 
follows when the temperature is low enough so that  H$_3^+$ + HD deuterium 
exchange reaction may proceed on the exothermic pathway as first proposed 
by \citet{watson74}. 
The H$_2$D$^+$ thus produced can further transfer its deuteron to the same 
neutral molecules, producing DCO$^+$, N$_2$D$^+$, at a rate 3 times smaller 
than in  the reaction involving H$_3^+$, as 2/3 of the reaction rate 
coefficient 
will lead also to HCO$^+$ and N$_2$H$^+$, if statistical arguments may 
apply. In the 
reaction of H$_2$D$^+$ with HD, D$_2$H$^+$ is formed, and a further reaction 
with HD  then leads to the completely deuterated ion D$_3^+$. These processes 
have received attention both in astrophysics 
\citep{roueff00,roberts00,rob03,fl:04}, and in 
chemical physics studies \citep{gerlich02,gers02}. \citet{rt:04} have summarized 
the exothermicities involved in the 
possible reactions involving deuterated substitutes of H$_3^+$ and H$_2$, 
which are typically between 150K and 220K. 
Temperatures as those derived in Section \ref{clouds} are thus low 
enough to inhibit the occurrence of the corresponding reverse reactions 
and suitable for deuterium enhancement to occur.
The confirmed detections of H$_2$D$^+$ in the pre-stellar core LDN~1544 
\citep{caselli03, vastel06}, as well as the recent detection of  
D$_2$H$^+$  in LDN~1689N  \citep{vastel04} , have provided strong
observational support for this theory. 
 It is important to notice that the isotopologue ions thus formed 
(N$_2$H$^+$, N$_2$D$^+$  on the one hand, HCO$^+$, DCO$^+$  on the other hand) 
are then destroyed by the same chemical reactions, i.e., dissociative 
recombination, reactions with the other abundant neutral species, 
possible charge transfer reactions with metals, recombination on 
grains, ...  In the absence of specific measurements, one assumes 
that the rate coefficients involving hydrogenated and deuterated 
molecular ions are equal. The steady state fractionation ratios
  N$_2$D$^+$  /  N$_2$H$^+$,  DCO$^+$ / HCO$^+$  are  then directly equal  
to the ratio of the formation probabilities of  N$_2$H$^+$ and N$_2$D$^+$ 
on the one hand and HCO$^+$ and DCO$^+$ on the other hand.
  
 In the case of N$_2$H$^+$, the main formation route is usually  
N$_2$ + H$_3^+$,  whereas for N$_2$D$^+$, the formation channels 
involve N$_2$ + H$_2$D$^+$,  N$_2$ + D$_2$H$^+$ and N$_2$ + D$_3^+$ . N$_2$H$^+$ 
may also be produced via N$_2$ + H$_2$D$^+$ and  N$_2$ + D$_2$H$^+$ if 
these ions become abundant. At steady state, the abundance ratio is given by : 
\begin{eqnarray}
\frac{\textrm{X}(\textrm{N}_2\textrm{D}^+)}{\textrm{X}(\textrm{N}_2\textrm{H}^+)} 
=    
\frac{
\textrm{X}(\textrm{H}_2\textrm{D}^+) 
+ 2 \, \textrm{X}(\textrm{H}\textrm{D}_2^+) 
+ 3 \, \textrm{X}(\textrm{D}_3^+)}
{ 
  \textrm{X}(\textrm{H}\textrm{D}_2^+)
+  2 \, \textrm{X}(\textrm{H}_2\textrm{D}^+)
+  3 \, \textrm{X}(\textrm{H}_3^+) }
\label{statistical}
\end{eqnarray}
where X(x)  represents the abundance or the fractional abundance of 
a particular species x. This relation holds if the overall reaction 
rate coefficients of N$_2$ with the various deuterated isotopologues 
of H$_3^+$ are equal and if statistical equilibrium determines  
the branching ratios.
 
 High fractionation ratios such as those found in the studied 
environments are then directly correlated to high deuteration 
fractions of the deuterated H$_3^+$  ions which can occur in 
highly depleted environments \citep{rob04,walmsley04,fl:04,roueff05}.
In addition to these reactions, N$_2$D$^+$ (DCO$^+$)  may also be 
formed in the deuteron exchange reaction between  D atoms 
and N$_2$H$^+$ (HCO$^+$).

Density and temperatures are derived from the analysis of 
the line profiles as shown in Table \ref{table:models}. The deduced fractional 
abundances of  N$_2$H$^+$, which are of the order of 10$^{-10}$ 
compared to H$_2$, imply a first constraint on the depletions, 
i.e. the available abundances of the elements in the gas phase. 
The deuterium fractionation ratio brings an additional limit on 
these values and also on other physical parameters such as the cosmic ray ionization 
rate ($\zeta$) which is directly related to the ionization fraction.

\begin{figure}
\centering
\includegraphics[scale=0.75]{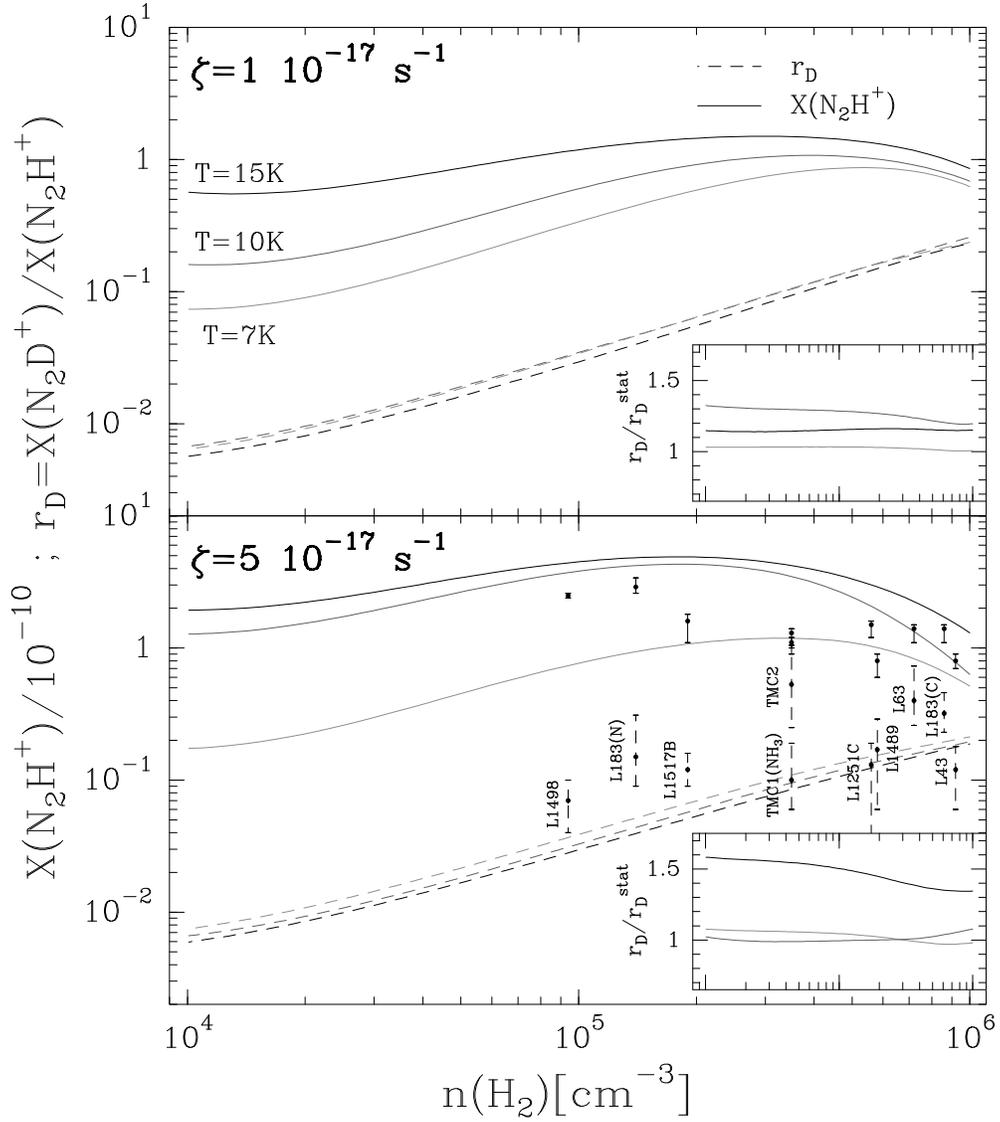}
\caption{N$_2$H$^+$ fractional abundances (solid lines, in unit of 10$^{-10}$) 
and ratio of abundances $r_D$=X(N$_2$D$^+$)/X(N$_2$H$^+$) (dashed lines), for a cosmic ray 
constant of $\zeta=1 \, 10^{-17}$ s$^{-1}$ (upper panel) and 
$\zeta=5 \, 10^{-17}$ s$^{-1}$ (lower panel). For each panel, the results 
are shown for clouds at T = 7K (faint grey), T = 10K (grey) and T = 15K (black).
The lower--right box of each panel compares the ratio $r_D$
to the ratio $r_D^{stat}$ predicted using equation \ref{statistical}.
Central densities (n$_0$) ,
N$_2$H$^+$ abundances, and abundance ratio ($R$) derived from models (see section \ref{clouds})
and listed in tables \ref{table:models} and \ref{table:models2} are indicated in the lower panel, for comparison
with the chemical models.}
\label{graph:chimie}
\end{figure}

Steady state model calculations allow to span rapidly the parameters 
space and to obtain the main physical and chemical characteristics of 
the environment. We have used an updated  chemical model described in 
\citet{roueff05}, where we include multiple deuterated species containing 
up to 5 deuterium atoms (CD$_5^+$) and the most recent chemical reaction rate 
coefficients. The gas phase chemical network includes 210 species 
containing H, D, He, C, N, O, S and a typical metal undergoing charge 
transfer reactions which are connected through about 3000 chemical 
reactions. Particular care is paid to 
dissociative recombination branching ratios\footnote{In the current chemical analysis, 
the branching ratios of 
dissociative recombination of N$_2$H$^+$ are taken from \citet{gep04}.
During the revision of the manuscript, it has been brought to our attention that in 
in a recent experiment, \citet{mol07} have reanalysed these  
branching ratios and obtained a  
major channel of reaction towards N$_2$, opposite to what 
\citet{gep04} had found with the storage ring experiment. 
For some test cases, 
we have introduced these new branching ratios, by keeping the same value of  
the total destruction rate of N$_2$H$^+$ and N$_2$D$^+$ and 
we have found that the fractional abundances of N$_2$H$^+$, N$_2$D$^+$ as 
well as the fractionation ratio remain very similar to the one displayed on figure \ref{graph:chimie}. 
However, substantial changes are found for hydrogenated nitrogen molecules and their 
isotopologues such as NH, NH$_2$, ...}
which modify considerably the chemical composition 
of the gas. 
We mimic the various mechanisms involved in gas-grain interactions
(accretion, thermal desorption, cosmic-ray induced desorption,
etc ...) by introducing a dependence of the available gas phase
elemental abundances with density.
The variation laws assumed for the gas phase elemental abundances
with density are arbitrary and account for the decrease of gas phase
atomic carbon, oxygen, nitrogen and sulfur with increasing density.
The assumed laws are:

\begin{eqnarray}
    n_C/n_H & = & 1.40 \, 10{^{-4}} \left[ 1 - 
    \textrm{exp} \left(- 4 \, 10^4 / n_H \right) \right]  \nonumber \\
    n_N/n_H & = & 7.94 \, 10{^{-5}} \left[ 1 - 
    \textrm{exp} \left(- 3 \, 10^4 / n_H  \right) \right] \nonumber \\ 
    n_O/n_H & = & 3.50 \, 10{^{-4}} \left[ 1 - 
    \textrm{exp} \left(- 4 \, 10^4 / n_H \right) \right] \nonumber \\
    n_S/n_H & = & 1.85 \, 10{^{-5}} \left[ 1 - 
    \textrm{exp} \left(- 4 \, 10^4 / n_H \right) \right]  \nonumber
\end{eqnarray}
where $n_x$ stands for the elemental volume density of species $x$. 
At low densities, the values correspond to the abundances observed in 
diffuse and translucent environments.
The carbon to oxygen ratio is fixed throughout the density variations.
For H$_2$ 
densities in the range between 10$^5$ and 10$^6$ cm$^{-3}$, the 
elemental abundances of C, N, O and S are in the range of several 
10$^{-5}$ - 10$^{-6}$, corresponding to depletions of 10 - 100. 

We solve the charge balance on the grains and allow the atomic ions to 
neutralize on the mostly negatively charged grains. These processes are 
important for the overall ionization fraction of the gas and lead to 
very low ionization fraction of the order of 10$^{-9}$ for densities larger 
than 10$^5$ cm$^{-3}$. Such low values are compatible with the derived 
value by \cite{cas02b} from observed deuterium fractionation in dense cores.
We show in Fig. \ref{graph:chimie}, both the fractional abundance of N$_2$H$^+$ 
in units of 10$^{-10}$ and the N$_2$D$^+$  / N$_2$H$^+$ ratio for densities ranging 
from 10$^4$ to 10$^6$ cm$^{-3}$ and a 
temperature of 10K. Two $\zeta$ values are used 
to probe the role of this important physical parameter.

We see that the obtained results span very nicely the values derived 
from the observations. To model the N$_2$H$^+$ emission, it has been assumed that 
the density structure
was consisting on a power law. We thus derived that observed spectra 
were consistent with an increase of the temperature with radius, and that,
depending on the source, the N$_2$H$^+$ abundance was whether constant or 
decreasing outward.
Considering the two $\zeta$ values assumed we find that the trend derived from the modeling is 
qualitatively reproduced with $\zeta = 1 \, 10^{-17}$ s$^{-1}$. 
On the other hand, for a higher $\zeta$ value, the abundance
would tend to increase with radius.

The models done in section \ref{clouds} show that the N$_2$D$^+$/N$_2$H$^+$ ratio 
could reach values as high as 0.5 in the innermost regions of the clouds.
Figure \ref{graph:chimie} shows that chemistry predict a 
N$_2$D$^+$/N$_2$H$^+$ ratio greater than 0.1 for densities 
higher than 2--3 $10^5$ cm$^{-3}$, with maximum values of 0.2--0.3. 
While the maximum values derived 
from the chemical modeling and from observations are within a factor 2,
the trend derived for the ratio, in both cases, are in good agreement.
Actually, for four sources (L63, TMC2, and the two dense cores of L183),
we obtained from observations that the
ratio should decrease rapidly with density, a trend reproduced
in the chemical modeling.   

Note that both absolute abundances and abundances ratio have been
 found 
to be sensitive to the adopted depletion law. In contrast, 
Figure \ref{graph:chimie} shows that the N$_2$D$^+$/N$_2$H$^+$ 
ratio is rather insensitive to variations of the temperature 
or cosmic ray constant, while the N$_2$H$^+$ and N$_2$D$^+$ absolute 
abundances are strongly changed.
We consider here only gas phase processes so that the role of induced
desorption by cosmic rays is assumed to be constant within the
variation by a factor of 5 of the cosmic ray ionization rate value.
A variation of $\zeta$ changes both the abundances of H$_3^+$
isotopologues and of the electrons, which are involved respectively
in the main N$_2$H$^+$ and N$_2$D$^+$ formation and destruction routes. 
The larger efficiency of formation from H$_3^+$ isotopologues entails 
that the absolute abundances increase with $\zeta$. Moreover, the similarity
of the formation and destruction channels of N$_2$H$^+$ and N$_2$D$^+$
lead to nearly equal variations, both with the temperature and $\zeta$, which
explains why the isotopologues abundances ratio remain identical. 
Finally, we point out that the N$_2$D$^+$/N$_2$H$^+$ ratio could serve 
as a probe of the way the C,N,O and S containing species deplete onto grains, 
since other parameters such as temperature and the cosmic ray ionization rate just influence 
the isotopologues absolute abundances similarly.

Figure \ref{graph:chimie} shows the N$_2$D$^+$/N$_2$H$^+$ ratio derived from
the full chemical network in comparison to the ratio expected from
statistical considerations (i.e. equation \ref{statistical}). We find a 
reasonable agreement between the two ratio, the largest differences (for T = 15K
and $\zeta = 5 \, 10^{-17}$ s$^{-1}$ ) being of the order of 60\%. 
The largest values in the full modeling are due to an increased 
N$_2$D$^+$ formation through the deuterium exchange reaction 
between D atoms and N$_2$H$^+$. For some specific conditions, this formation route 
is found efficient.

\section{Conclusions}\label{conclusion}
We have derived estimates of the temperature, density,
N$_2$H$^+$ and N$_2$D$^+$ abundances in a sample of cold dark clouds, by using a radiative transfer 
modeling which enables to interpret the hyperfine transitions of these molecules.
The main conclusions are :

\begin{enumerate}

\item Compared to previous studies considering N$_2$H$^+$, we generally derive 
higher densities due to the inclusion in the models of the $J$=2--1 line. For 
the conditions prevailing in dark clouds, i.e. T$_K$ $\sim$ 10 K, this line is the 
optically thickest and, in order to prevent strong self absorption effects which 
are generally not observed, models with high density and low abundance are
preferred. Moreover, an outward increase of the temperature enables to reduce the 
self absorption in this line.

\item In order to reproduce the $J$=1--0 hyperfine transitions, the total opacity of this line has to 
be larger than $\tau$ $\sim$ 10. A way to increase the opacity and to still reproduce 
the other rotational lines is to decrease the temperature in the inner core below 
T = 10 K. For most sources, we encounter a good agreement with a temperature around 
T = 8 K in the central region of the clouds.   

\item We analyzed the density structure of the clouds and we took into account 
previous studies where the $J$=1--0 integrated intensity, as a function of the position on
the cloud, were reported. 
The central average densities and the radii of the inner flat regions, derived using 
N$_2$H$^+$, are in good agreement with the equivalent parameters obtained
from sub--mm and mm continuum observations. Compared to other studies based 
on other molecular species, it seems that N$_2$H$^+$ is the only one which allows 
such a good agreement, since other molecules provide systematically lower H$_2$ volume 
densities. For most molecules, it still might be possible to conciliate 
these estimates by introducing the radial dependence in the molecular abundances which
are predicted by theoretical studies of dark clouds chemistry.

\item We derive X(N$_2$D$^+$)/X(N$_2$H$^+$) which are in qualitatively
good agreement with previous studies. Moreover, for two of the studied 
clouds (TMC2 and L183 ) where we have observed various positions in the cloud, 
we find that the observed spectra are best reproduced when increasing the D/H ratio in the 
inner dense regions. We note that for these 2 objects a single position 
analysis lead to the derivation of a constant ratio throughout the cloud, which was 
smaller than the central value obtained in the multi position analysis. Thus, for the other 
objects, the ratio we have derived has to be considered as an average value and a central 
enhancement of the D/H ratio cannot be ruled out. 

Considering studies that deal with the chemical evolution of 
protostellar clouds, our results are in qualitative and quantitative 
agreement with the expected trend of the two isotopologues \citep{aik05}. Recently, the 
importance of the multideuterated species of H$_3^{+}$ in the chemical network was pointed out 
by \citet{rob03} in order to explain the high deuterium fractionation observed 
in molecular clouds. This work shows that for various molecular species
we should observe D/H ratios approaching unity. Moreover, they show that the deuterium fractionation 
is particularly efficient in the case of N$_2$H$^+$. A detailed time dependent study 
considering both dynamics and chemistry was reported in \citet{aik05} 
for the stages prior to star formation. It confirms that deuterium fractionation can 
reach high values for N$_2$H$^+$ and that the abundance ratio takes larger values in the 
innermost regions of the clouds. From this latter study, it appears that the fractionation 
is expected to increase with time when the central density of the inner region increases. 
Thus, the determination of the deuterium enrichment could serve as a tool to probe 
the stage of evolution prior to the formation of protostars.

\end{enumerate}

\acknowledgments
The authors thank P. Caselli for providing N$_2$H$^+$ observational data.
The authors are grateful to I. Jim\'enez--Serra, J.--R. Pardo and M. Ag\'undez for their
help and suggestions concerning the present manuscript.
The authors acknowledge funding support from the CNRS/INSU programme PCMI,
and from the European Union Marie-Curie network "Molecular Universe".
J. Cernicharo wants to thank Spanish MEC for funding support
under grants PANAYA2000--1784, ESP2001--4516, PNAYA2002--10113--E, ESP2002--01627, 
PNAYA2003--02785, PNAYA2004--0579  and the Madrid DGU PRICIT program S-0505/ESP--0237
(ASTROCAM).

{\bf{Appendix: Radiative Transfer Models for N$_2$D$^+$}}

The rotational and hyperfine coupling constants for N$_2$D$^+$
are taken from \citet{dor04}.
We have used the dipole moment of N$_2$H$^+$, i.e. $\mu$ = 3.4 D, to derive 
the line strengths of the N$_2$D$^+$ hyperfine lines, according to equation (2) of 
\citetalias{dan06}. For the hyperfine collisional rate coefficients, we have used the 
de--excitation rates of N$_2$H$^+$ colliding with He \citep{dan05} and applied 
the detailed balance relationships to derive the excitation rate
coefficients (see below).

Figure \ref{graph:LVG_isotope} shows the excitation temperatures, 
opacities and brightness temperatures of the $J$=1--0, 2--1 and 3--2 transitions 
of N$_2$H$^+$ and N$_2$D$^+$ obtained under the LVG approximation. The column 
density is the same for both species and equal to 
10$^{12}$ cm$^{-2}$ / (km s$^{-1}$ pc$^{-1}$).
In order to derive the rotational line opacities we have to
take into account the hyperfine structure \citepalias[see][]{dan06}.
Excitation temperatures are obtained according to equation (7)
of \citetalias{dan06}.

The difference in the excitation processes 
for the two isotopologues is a consequence of their specific rotational 
energy structure, and hence, different frequencies and
Einstein coefficients for the rotational lines.
For N$_2$D$^+$, the frequencies of the rotational transitions are lower
than those of the main isotopologue by a factor B(N$_2$H$^+$)/B(N$_2$D$^+$) = 1.21. 
Consequently, the Einstein coefficients
are lower for N$_2$D$^+$ by a factor 1.8, which results in lower critical
densities if we assume that the de--excitation rate
coefficients are identical for both species.
This latter assumption has been checked by comparing the rotational rate coefficients of 
N$_2$H$^+$ and N$_2$D$^+$ colliding with He (see table \ref{table:rate_N2D+}).
The N$_2$D$^+$ rate coefficients have been computed by the
MOLSCAT\footnote{J. M.
Hutson and S. Green, MOLSCAT computer code, version 14 (1994), distributed by 
Collaborative Computational Project No. 6 of the Engineering and Physical Sciences 
Research Council (UK).} code using the potential energy surface given by 
\citet{dan04}. The reduced mass of the colliding system has been
modified with respect to the N$_2$H$^+$ calculations. However, we have
assumed that the mass center for both isotopologues were the same (this
enables a straightforward determination of rate coefficients for the
deuterated species).
The results are given in Table \ref{table:rate_N2D+}. We can see that for
temperatures ranging from 10 to 30 K, the differences between the rotational
de--excitation rate coefficients of the two isotopologues are $<$ 20\%. 
However, upwards rate coefficients are obtained from detailed balance and
they will be different for both isotopologues, being larger for the deuterated
species. The difference is small, of the order of 10--20\%,
for $\Delta J$=+1 transitions, but becomes as large as a factor 4
for higher $\Delta J$.
The lower Einstein A coefficients and higher
upwards rate coefficients for N$_2$D$^+$ with respect to N$_2$H$^+$ allow 
the high--J rotational levels to be more efficiently populated for the deuterated species.
Hence, the effect introduced by the different rotational constants has
to be taken into account when computing the level population for both
isotopologues.
Figure \ref{graph:LVG_isotope} shows that for identical column densities
(10$^{12}$ cm$^{-2}$)
the $J$=1--0 and 3--2 lines of N$_2$D$^+$
have, respectively, lower and higher opacities than those of N$_2$H$^+$.
This behavior applies to the whole range of
densities explored in our calculations.
Nevertheless, the change in the critical density is only apparent for
the $J$=1--0 line (see T$_{ex}$ panel of Figure \ref{graph:LVG_isotope}).
The effect is less evident for the other transitions because
rotational excitation temperatures are obtained from the average of the
individual hyperfine T$_{ex}$ \citepalias[c.f.][]{dan06}.

The expected variation of critical densities is more obvious when
considering HCO$^+$,
H$^{13}$CO$^+$ and DCO$^+$. Figure 
\ref{graph:LVG_isotope} (right panels) shows that the opacities
for each transition vary similarly than N$_2$H$^+$. Moreover, 
on the T$_{ex}$ panel of Figure
\ref{graph:LVG_isotope} we see that the critical
densities are lower for DCO$^+$ and H$^{13}$CO$^+$ than for the main
isotopologue.  It is worth noting that the intensity ratios
T$_B$(DCO$^+$)/T$_B$(HCO$^+$) and T$_B$(N$_2$D$^+$)/T$_B$(N$_2$H$^+$),
depend on the volume density  for 10$^4$ $<$ n(H$_2$) $<$ 10$^6$
cm$^{-3}$ (see bottom panels of Figure \ref{graph:LVG_isotope}).
This behavior is found even for higher column densities.
If the column densities for the deuterated and main species are not the
same, then the plots in the bottom panels of Figure \ref{graph:LVG_isotope}
should be scaled by the assumed column density ratio.

The effect discussed above for singly deuterated linear molecules
could be even more important for multiply deuterated
species. In order to evaluate qualitatively this effect we have
considered the excitation of two asymmetric top molecules, 
o--H$_2$CO and p--D$_2$CO (note that the quantum numbers for ortho--formaldeyde 
and its doubly para--deuterated isotopoloque
are the same). Figure \ref{graph:formaldehyde} 
shows the excitation temperature in the optically thin case 
(column density of 10$^{12}$ cm$^{-2}$ for both isotopologues)
of different rotational transitions of these species 
as well as the corresponding line intensity ratios. 
The difference in the Einstein
coefficients for p--D$_2$CO and o--H$_2$CO, and the larger
upwards excitation rates for the former, produce an
important variation in the excitation conditions of
these two molecules. The line intensity ratio varies
from 1.7 for the transition 2$_{12}$-1$_{11}$ to 3.4
for the 4$_{14}$-3$_{13}$ one. We stress that this
difference in the excitation conditions
will translate into different emitting volumes
for each species.
Hence, the determination of the deuteration
degree
can not be directly related to the column
density ratios of the main and deuterated isotopologues. The
cloud density and temperature structure has to be used, together with
non-local radiative transfer codes, in order to derive from observations
the deuteration degree through the cloud.

The effect will be even more important for triply deuterated
species like CD$_3$OH or CD$_3$CN for which the rotational constant $A$
will change by a factor $\simeq$2.0 and $B$ and $C$ by a factor
$\simeq$1.5 \citep[see e.g.][]{wal98}. Thus, 
the energies of the CD$_3$OH levels are lower by a factor 
$\simeq$ 2 compared to CH$_3$OH. Hence, from detailed balance
relationships, upward collisional rates will be found very
different for both isotopologues. The global effect in the
determination of the deuteration enrichment will also
depend on the source physical structure. While the high
observed beam averaged abundance ratios for CH$_3$OH and its doubly
and triply deuterated isotopologues can not be ruled out by these
excitation effects
\citep[see e.g.][]{par02,par04}, an exact determination of these ratios 
should take into account the different cloud emitting volumes
for these species. The implication for chemical models is obvious
since a direct comparison of the derived column densities
will always produce higher deuteration factors than the real ones.
Thus, detailed radiative transfer models have to be carried out to derive
correct values for the deuteration enrichment.

\clearpage

\begin{table}
\begin{tabular}{|c|c|c|c|c|c|}
\hline
Source & $\alpha$(J2000) &  $\delta$(J2000) & I(K km s$^{-1}$) & r(pc) & D(pc) \\
\hline
L1489            & 04 04 49.0 &  26 18 42 & 1.8 & 0.039 & 140  \\ 
L1498            & 04 10 51.4 &  25 09 58 & 1.5 & 0.051 & 140  \\ %
TMC2             & 04 32 46.8 &  24 25 35 & 2.5 & 0.061 & 140  \\ 
TMC1(NH$_3$)$^1$ & 04 41 21.3 &  25 48 07 & 2.3 & 0.040 & 140  \\ %
L1517B           & 04 55 18.8 &  30 38 04 & 1.4 & 0.050 & 140  \\ %
L183(C)          & 15 54 08.7 & -02 52 07 & 2.5 & 0.045 & 110  \\ %
L183(N)          & 15 54 09.2 & -02 49 39 & 1.8 & 0.045 & 110  \\ %
L43              & 16 34 35.0 & -15 46 36 & 3.4 & 0.059 & 160  \\ 
L63              & 16 50 15.5 & -18 06 26 & 2.0 & 0.057 & 160  \\ 
L1251C           & 22 35 53.6 &  75 18 55 & 1.7 & 0.069 & 200  \\ 
\hline
\end{tabular}
\caption{ Right ascension and declination of the observed sources, assumed distance (D), 
integrated intensity of the $J$=1--0 line (I)
and half power radius of the models (r), as obtained by convolution with the FCRAO antenna beam,
for comparison with the observed values given in \citet{cas02}.
\newline
$^1$ For this source, the half power radius and the central intensity are calculated considering
only the main emission component.}\label{table:obs_mod}
\end{table}
\clearpage
\renewcommand{\arraystretch}{1.3}
\begin{sidewaystable}
\scriptsize
\begin{tabular}{|l||r|r||r|r|r||r|r|r||r|r|r|}
\hline
Source & r$_0$["] &  n$_0$$/10^5$ [cm$^{-3}$] & r$_1$["] & T$_1$[K] & X$_1$$/10^{-10}$ & r$_2$["] & T$_2$[K] & X$_2$$/10^{-10}$ & r$_3$["] & T$_3$[K] & X$_3$$/10^{-10}$  \\
\hline
L1251C       & 29$^{+1}_{-3}$ & 5.6$^{+0.1}_{-0.5}$  &  21 & 7$^{+0.2}_{-0.4}$   & 1.5$^{+0.1}_{-0.3}$  & 150 & 10$^{+1.1}_{-2.3}$  & 0.2$^{+0.1}_{-0.1}$ & ... & .. & ...  \\ 
L43          & 29$^{+1}_{-1}$ & 9.2$^{+0.4}_{-0.3}$  &  30 & 8$^{+0.3}_{-0.9}$   & 0.8$^{+0.1}_{-0.1}$  & 60  & 11$^{+1.8}_{-2.0}$  & 0.4$^{+0.1}_{-0.1}$ & 150 & 14$^{+8}_{-3}$ & 0.3$^{+0.1}_{-0.1}$  \\ 
TMC2         & 60$^{+4}_{-3}$ & 3.5$^{+0.1}_{-0.1}$  &  40 & 8$^{+0.2}_{-0.6}$  & 1.1$^{+0.1}_{-0.2}$  & 200 & 10$^{+1.8}_{-1.4}$  & 0.2$^{+0.1}_{-0.1}$ & ... & ... & ...  \\ 
L63          & 23$^{+1}_{-1}$ & 7.2$^{+0.4}_{-0.5}$  &  20 & 8$^{+0.3}_{-0.6}$   & 1.4$^{+0.1}_{-0.3}$  & 45  & 8$^{+0.8}_{-2.0}$   & 0.4$^{+0.1}_{-0.1}$ & 150 & 15$^{+7}_{-4}$ & 0.4$^{+0.2}_{-0.2}$  \\ 
L1489        & 24$^{+1}_{-1}$ & 5.8$^{+0.3}_{-0.4}$  &  35 & 8$^{+0.6}_{-0.7}$   & 0.8$^{+0.1}_{-0.2}$  & 65  & 9$^{+2.8}_{-1.6}$   & 0.8$^{+0.4}_{-0.1}$ & 100 & ? & 0.1$^{+0.2}_{-0.1}$  \\
L1498        & 70$^{+7}_{-5}$ & 0.94$^{+0.03}_{-0.02}$ &  70 & 8$^{+0.3}_{-0.2}$   & 2.5$^{+0.1}_{-0.1}$  & 200 & 10$^{+5}_{-2}$  & 0.3$^{+0.1}_{-0.1}$ & ... & .. & ...  \\ 
TMC1(NH$_3$) & 30$^{+1}_{-1}$ & 3.5$^{+0.1}_{-0.1}$  &  35 & 8$^{+0.4}_{-1.6}$   & 1.3$^{+0.1}_{-0.3}$  & 65  & 12$^{+1.8}_{-1.6}$  & 1.3$^{+0.2}_{-0.1}$ & 175 & ? & 0.1$^{+0.6}_{-0.1}$  \\ 
L1517B (1)   & 30$^{+1}_{-1}$ & 1.9$^{+0.1}_{-0.1}$  &  18 & 8.5$^{+1.3}_{-1.1}$ & 1.6$^{+0.2}_{-0.5}$  & 180 & 9.5$^{+0.2}_{-0.3}$ & 1.6$^{+0.1}_{-0.1}$ & ... & .. & ...  \\ 
L1517B (2)   & 35$^{+1}_{-1}$ & 2.0$^{+0.1}_{-0.1}$  &  18 & 8.5$^{+2.0}_{-1.0}$ & 2.5$^{+0.7}_{-1.2}$  & 180 & 9.5$^{+0.5}_{-0.3}$ & 2.5$^{+0.2}_{-0.1}$ & ... & .. & ...  \\ 
L183(C)      & 19$^{+1}_{-1}$ & 8.6$^{+0.2}_{-0.6}$  &  20 & 8$^{+0.4}_{-0.6}$   & 1.4$^{+0.1}_{-0.3}$  & 40  & 9$^{+0.6}_{-1.6}$   & 1.4$^{+0.1}_{-0.6}$ & 225 &  9$^{+0.5}_{-3}$ & 2.1$^{+0.2}_{-0.3}$  \\ 
L183(N)      & 44$^{+4}_{-4}$ & 1.4$^{+0.1}_{-0.1}$  &  80 & 9$^{+1.1}_{-0.8}$   & 2.9$^{+0.5}_{-0.3}$  & 225 & 9$^{+9}_{-4}$   & 2.0$^{+4.0}_{-0.6}$ & ... & .. & ... \\ 
\hline
\end{tabular}
\caption{
Parameters obtained from the modeling of our observations of N$_2$H$^+$
lines for the 10 clouds studied. The density profile is described as n(H$_2$)
= n$_0$ for r $<$ r$_0$ and n(H$_2$) = n$_0$(r$_0$/r)$^2$ for r $>$ r$_0$.
Depending on the source, we have introduced 2 or 3 components  
to describe the cloud: the radius of the i$^{th}$ component is given as r$_i$ 
and for radii $r$ in the range $r_{i-1} < r < r_{i}$, the N$_2$H$^+$ abundance and gas temperature
are respectively given as X$_i$ and T$_i$ (the first zone with T$_1$ and X$_1$, corresponds to the radius range $0 < r < r_1"$). For two clouds, i.e. L1489 and TMC1(NH3), the temperature 
of the $3^{rd}$ zone is poorly constrained by the available observations. The large error bars found 
make the value derived non significant and this parameter is thus not indicated.}
\label{table:models}
\end{sidewaystable}

\begin{sidewaystable}
\scriptsize
\begin{tabular}{|l|c|c|c|c|c|}
\hline
Source &  M(r<2') [M$_\sun$] & r$_D$ & N(N$_2$H$^+$)$/10^{12}$ [cm$^{-2}$] & N(N$_2$D$^+$)$/10^{12}$ [cm$^{-2}$] & R \\
\hline
L1251C    & 31.1$^{+2.0}_{-5.6}$ & 0.14$^{+0.03}_{-0.11}$     & 12.6$^{+1.1}_{-2.1}$ & 1.6$^{+0.4}_{-1.3}$ & 0.13$^{+0.06}_{-0.11}$ \\ 
L43          & 26.2$^{+1.6}_{-1.6}$ & 0.12$^{+0.04}_{-0.06}$     & 14.1$^{+1.3}_{-1.3}$ & 1.7$^{+0.6}_{-0.8}$ & 0.12$^{+0.06}_{-0.06}$ \\ 
TMC2       & 22.7$^{+2.3}_{-1.7}$ & 0.73$^{+0.53}_{-0.38}$$^a$ & 8.2$^{+0.9}_{-1.2}$ & 4.4$^{+3.0}_{-2.1}$ &  0.53$^{+0.52}_{-0.28}$ \\ 
L63          & 13.4$^{+1.1}_{-1.1}$ & 0.64$^{+0.56}_{-0.34}$$^b$ & 12.7$^{+0.7}_{-2.0}$ & 5.1$^{+2.7}_{-1.6}$ & 0.40$^{+0.33}_{-0.14}$ \\ 
L1489        & 7.8$^{+0.6}_{-0.6}$  & 0.17$^{+0.06}_{-0.11}$     & 7.7$^{+0.8}_{-1.5}$  & 1.3$^{+0.5}_{-0.8}$ & 0.17$^{+0.12}_{-0.11}$ \\
L1498         & 7.6$^{+1.0}_{-0.8}$  & 0.07$^{+0.02}_{-0.02}$     & 7.3$^{+0.3}_{-0.2}$  & 0.5$^{+0.2}_{-0.1}$ & 0.07$^{+0.02}_{-0.02}$ \\
TMC1(NH$_3$)  & 7.1$^{+0.4}_{-0.4}$  & 0.10$^{+0.07}_{-0.03}$     & 8.9$^{+0.8}_{-1.0}$  & 0.9$^{+0.6}_{-0.3}$ & 0.10$^{+0.09}_{-0.04}$ \\ 
L1517B (1)    & 3.9$^{+0.2}_{-0.3}$  & 0.12$^{+0.03}_{-0.02}$     &  6.3$^{+0.4}_{-0.4}$ & 0.8$^{+0.2}_{-0.2}$ & 0.12$^{+0.04}_{-0.03}$ \\
L1517B (2)   & ...  & 0.10$^{+0.03}_{-0.02}$     & ..... & .... & .... \\ 
L183(C)       & 3.6$^{+0.4}_{-0.3}$  & 0.43$^{+0.05}_{-0.12}$$^c$ & 13.9$^{+0.4}_{-2.0}$ & 4.2$^{+1.6}_{-1.1}$ & 0.33$^{+0.16}_{-0.11}$ \\ 
L183(N)     & 2.7$^{+0.4}_{-0.4}$  & 0.76$^{+1.00}_{-0.20}$$^d$ & 9.9$^{+2.9}_{-0.8}$  & 1.4$^{+1.4}_{-0.2}$ & 0.15$^{+0.16}_{-0.06}$ \\ 
\hline
\end{tabular}
\caption{
For the clouds with model's parameters given in table \ref{table:models},
the mass enclosed within a radius of 2' is given (in M$_\sun$) as well as the 
abundance ratio r$_D$ = X(N$_2$D$^+$)/X(N$_2$H$^+$). The column densities
derived for the positions reported in table \ref{table:obs_mod} are quoted
and their ratio, R, is given in the last column.
\newline
$^a$ the quoted abundance ratio is for radii below 35'' and is
decreased to 0.10$^{+0.05}_{-0.10}$ for greater radii.
\newline
$^b$ the quoted abundance ratio is for radii below 10'' and
is decreased to 0.25$^{+0.16}_{-0.02}$ for greater radii.
\newline
$^c$ the quoted abundance ratio is for radii below 40'' and is
decreased to 0.05$^{+0.34}_{-0.05}$ for greater radii.
\newline
$^d$ the quoted abundance ratio is for radii below 10'' and is
decreased to 0.05$^{+0.02}_{-0.01}$ for greater radii.}\label{table:models2}
\end{sidewaystable}

\renewcommand{\arraystretch}{1.0}

\begin{table}
\begin{tabular}{|c|c|c|c|c|c|}
\hline
 & \multicolumn{3}{c}{N$_2$H$^+$} & N$_2$D$^+$ & \\ \hline
Source   & V$_{LSR}$ & $\delta$v$_{LSR}$&
$\delta$v$_{LSR}$& $\delta$ v$_{LSR}$& $\Delta$v\\
         & ($J$=1--0) & ($J$=2--1) & ($J$=3--2)& ($J$=3--2) & (km s$^{-1}$ pc$^{-1}$) \\
\hline
L1489    &  6.74 & 0.02 & 0.07 & 0.04 & -1.3 \\
TMC2     &  6.17 & 0.02 & 0.01 & 0.06 & -0.9 \\
L183(C)  &  2.40 & 0.01 & 0.04 & 0.04 & -0.2 \\
L183(N)  &  2.41 & .... & 0.01 & 0.06 & -0.0 \\
L43      &  0.74 & 0.03 & 0.04 & 0.01 & -0.4 \\
L63      &  5.73 & 0.02 & 0.04 & 0.03 & -0.0 \\
L1251C   & -4.74 & 0.02 & 0.02 & 0.02 & -0.5 \\
L1498    &  7.80 & .... & .... & .... & -0.0 \\
TMC1-NH3 &  5.98 & .... & .... & .... & -0.8 \\
L1517B   &  5.78 & 0.04 & .... & .... &  1.4 \\
\hline
\end{tabular}
\caption{V$_{LSR}$ obtained for the different transitions of N$_2$H$^+$ and N$_2$D$^+$, by minimizing 
the chi square between observed and modeled lines. All the values are given in comparison to the V$_{LSR}$
of the $J$=1--0 line, i.e. V$_{LSR}$ = V$_{LSR}$($J$=1--0) +  $\delta$ v$_{LSR}$. $\Delta$ v corresponds 
to the velocity gradient used in the models. Note that L1517B is the only cloud with an outward
gradient. \label{table:VLSR}}
\end{table}

\begin{table}
\begin{tabular}{|c|c|c c|c c|c c|}
\hline
 & & \multicolumn{2}{c}{T=10K} & \multicolumn{2}{c}{T=20K} & \multicolumn{2}{c}{T=30K} \\
\hline
 j & j' &  R$_{j \to j'}$(N$_2$D$^+$) & $\delta$ (\%)  &R$_{j \to j'}$(N$_2$D$^+$) & $\delta$ (\%)  &R$_{j \to j'}$(N$_2$D$^+$) & $\delta$ (\%)  \\
\hline
 1 &  0 &    103.39 &      -10.6 &     91.96 &      -10.7 &     88.51 &       -9.4 \\
 2 &  0 &     58.20 &       -7.9 &     47.91 &       -6.4 &     42.09 &       -4.0 \\
 3 &  0 &     35.04 &       -4.7 &     29.15 &       -6.5 &     24.98 &       -9.3 \\
 4 &  0 &     19.81 &      -16.5 &     17.97 &      -10.4 &     16.44 &       -9.4 \\
 5 &  0 &     19.33 &      -13.0 &     18.65 &      -12.8 &     18.01 &      -12.2 \\
 6 &  0 &     21.78 &       -2.7 &     19.41 &       -5.2 &     17.84 &       -4.9 \\ \hline
 2 &  1 &    175.25 &        0.0 &    156.25 &       -2.4 &    145.50 &       -3.3 \\
 3 &  1 &    107.85 &        5.7 &     89.89 &       -1.1 &     78.69 &       -3.7 \\
 4 &  1 &     64.71 &       -7.0 &     59.22 &       -7.6 &     54.97 &       -8.4 \\
 5 &  1 &     61.24 &      -13.9 &     55.96 &      -13.8 &     52.23 &      -13.0 \\
 6 &  1 &     59.52 &        1.6 &     53.17 &       -3.2 &     48.58 &       -4.6 \\ \hline
 3 &  2 &    172.15 &       -5.8 &    155.22 &       -5.7 &    147.37 &       -5.4 \\
 4 &  2 &    125.74 &      -11.7 &    114.11 &      -12.9 &    106.30 &      -12.8 \\
 5 &  2 &    112.87 &       -2.3 &    100.39 &       -4.8 &     90.80 &       -6.3 \\
 6 &  2 &     79.06 &        8.0 &     73.04 &        2.1 &     67.56 &       -0.9 \\ \hline
 4 &  3 &    196.35 &       -2.3 &    182.35 &       -3.4 &    173.38 &       -3.9 \\
 5 &  3 &    127.69 &        3.0 &    119.03 &        0.2 &    111.16 &       -2.1 \\
 6 &  3 &     95.84 &        4.2 &     93.03 &        2.8 &     88.18 &        0.5 \\ \hline
 5 &  4 &    140.44 &       -2.2 &    141.22 &       -0.5 &    141.61 &       -0.6 \\
 6 &  4 &    110.55 &        3.4 &    108.73 &        1.8 &    104.69 &       -0.3 \\ \hline
 6 &  5 &    119.72 &       -3.2 &    122.05 &       -0.7 &    124.50 &        0.0 \\
\hline
\end{tabular}
\caption{ Rotational rate coefficients for N$_2$D$^+$ colliding with He, in units of 10$^{-12}$ cm$^3$ s$^{-1}$,
 and percentage difference with the N$_2$H$^+$--He de--excitation rate coefficients. \label{table:rate_N2D+}}
\end{table}

\clearpage

\clearpage

\begin{figure}
\centering
\includegraphics[scale=0.65,angle=0]{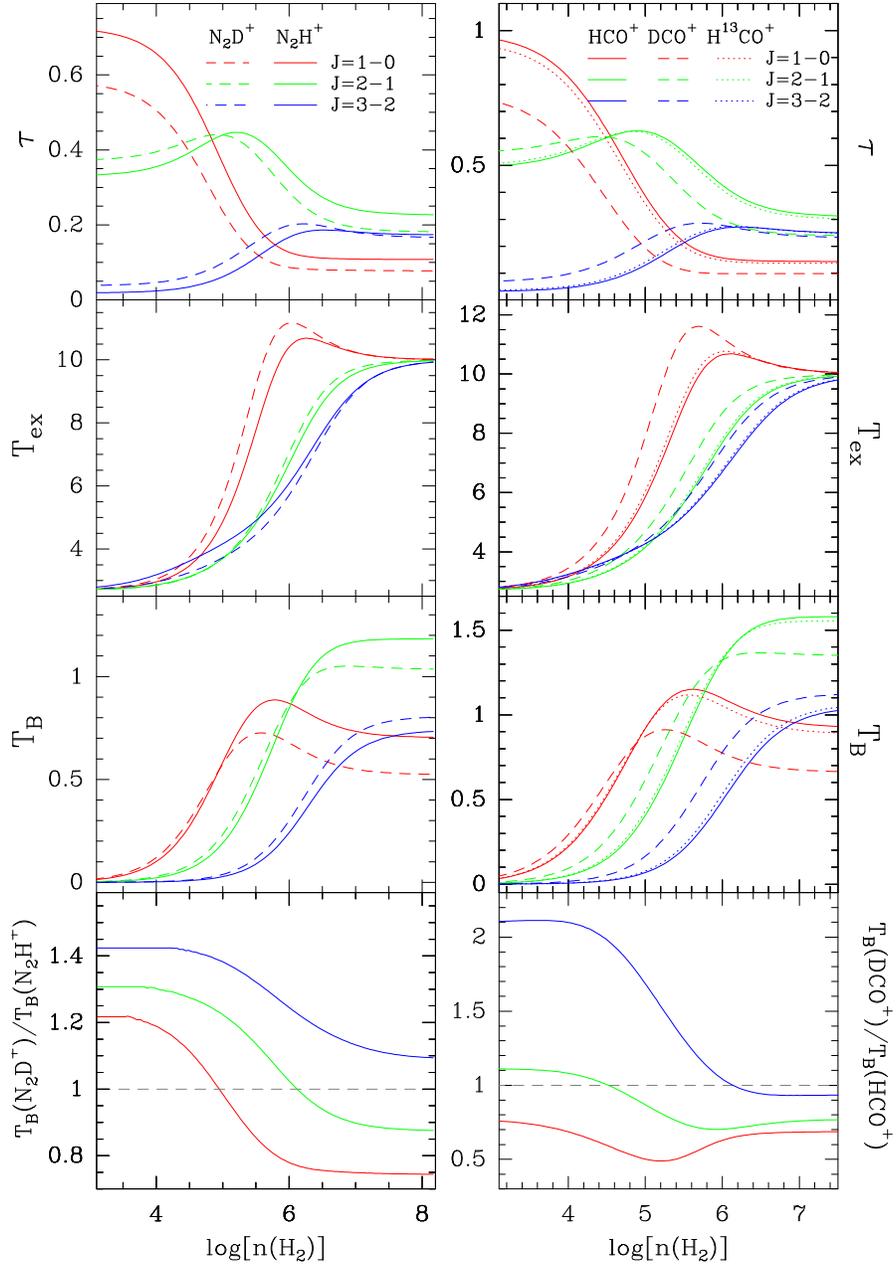}
\caption{$J$=1--0, 2--1 and 3--2 excitation temperature, opacity and brightness 
temperature of N$_2$D$^+$ (dashed lines) and N$_2$H$^+$ (solid lines) 
on the left column, and of HCO$^+$ (solid lines), H$^{13}$CO$^+$ (dotted line) 
and DCO$^+$ (long dashed) on the right column. The column density is the same 
for all molecules and is 10$^{12}$ cm$^{-2}$/(km s$^{-1}$ pc$^{-1}$).\label{graph:LVG_isotope}}
\end{figure}

\clearpage

\begin{figure}
\centering
\includegraphics[scale=0.65,angle=0]{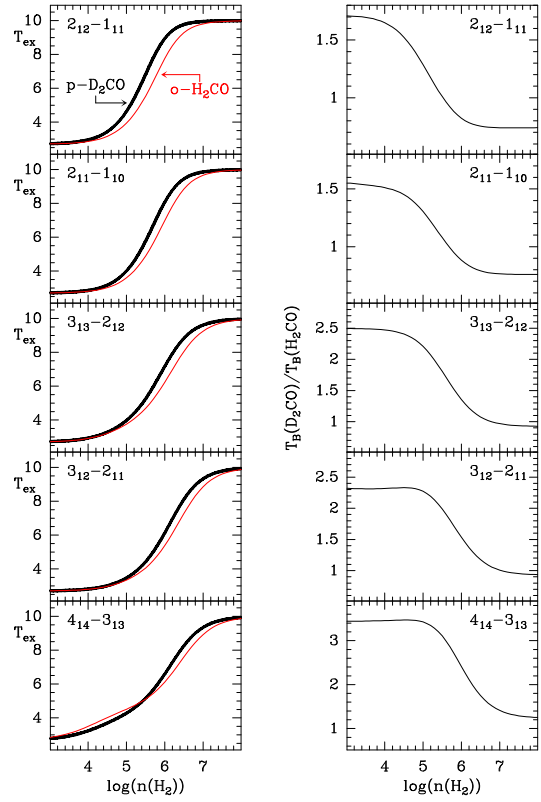}
\caption{ Comparison of excitation temperatures (left) and line
intensity ratios (rigth) of ortho--H$_2$CO and para--D$_2$CO for
different transitions of both species and a common column density
of 10$^{12}$ cm$^{-2}$.
Note that the quantum numbers of the selected transitions
correspond to different symetries of the two isotopologues. Left panels
show that excitation temperatures are different for both species
for the range of densities found in dense cores. 
\label{graph:formaldehyde}}
\end{figure}

\clearpage

\end{document}